\documentclass{elsart}
\journal{Advances in Water Resources}
\usepackage{graphicx}
\usepackage{amssymb}
\usepackage{amsmath}
\DeclareMathOperator{\erfc}{erfc}

\begin{document}
\begin{frontmatter}

\title{Groundwater age, life expectancy and transit time distributions
 in advective--dispersive systems:1. Generalized Reservoir Theory.}
\author{F. Cornaton\corauthref{cor}},
\corauth[cor]{Corresponding author.} \ead{fabien.cornaton@unine.ch}
\author{P. Perrochet}
\ead{pierre.perrochet@unine.ch}
\address{CHYN, Institute of Geology, University of Neuch\^{a}tel,
         Emile-Argand 11, CH-2007, Neuch\^{a}tel, Switzerland}

\begin{abstract} We present a methodology for determining reservoir
groundwater age and transit time probability distributions in a
deterministic manner, considering advective-dispersive transport in
steady velocity fields. In a first step, we propose to model the
statistical distribution of groundwater age at aquifer scale by
means of the classical advection-dispersion equation for a
conservative and non-reactive tracer, associated to proper boundary
conditions. The evaluated function corresponds to the density of
probability of the random variable age, age being defined as the
time elapsed since the water particles entered the aquifer. An
adjoint backward model is introduced to characterize the life
expectancy distribution, life expectancy being the time remaining
before leaving the aquifer. By convolution of these two
distributions, groundwater transit time distributions, from inlet to
outlet, are fully defined for the entire aquifer domain. In a second
step, an accurate and efficient method is introduced to simulate the
transit time distribution at discharge zones. By applying the
reservoir theory to advective--dispersive aquifer systems, we
demonstrate that the discharge zone transit time distribution can be
evaluated if the internal age probability distribution is known. The
reservoir theory also applies to internal life expectancy
probabilities yielding the recharge zone life expectancy
distribution. Internal groundwater volumes are finally identified
with respect to age and transit time. One- and two-dimensional
theoretical examples are presented to illustrate the proposed
mathematical models, and make inferences on the effect of aquifer
structure and macro--dispersion on the distributions of age, life
expectancy and transit time.
\end{abstract}
\end{frontmatter}
\section{Introduction.}



The knowledge of groundwater age distributions is of prime interest
in many environmental issues since they depend on the intrinsic
characteristics of the overall transport properties of an aquifer
and its sub-systems. For instance, an important fraction of young
water within a water sample can often be taken as the signature of a
reservoir with good turnover property. On the opposite, a
considerable component of old water may reflect a poorly recharged
aquifer, and/or significant internal mixing processes. The impact of
a contamination hazard on groundwater quality can be investigated
using groundwater age, since the age distribution provides direct
information on the time required for a water particle, or a
conservative tracer, to reach any critical zone that is to be
protected. The fate of a solute being transported in groundwater
partially depends on the time spent by the water molecules during
their migration throughout the aquifer system. Reactive transport of
a specific substance is also linked to groundwater age; the time
spent flowing through any kind of mineral heterogeneity being a
conditional factor for the development of any potential reactions.
With the age information, inferences can be made on the aquifer
physical characteristics, as well as on the chemical transformations
that contaminants may undergo. The age can also be of importance for
estimating historical aspects related e.g. to the agricultural
practices, or the land use of a particular region, which are
expected to have lead to groundwater contamination. As it is the
case throughout this work, groundwater age can be considered as a
fully conservative tracer. Consequently, the worst scenario with
regard to a contamination case is thus chosen by solving the age
transport problem.

Generally, a water sample shows a mixture of ages, which may range
between several orders of magnitude, as a consequence of the
reservoir geometrical and hydro-dispersive characteristics (spatial
repartition of the hydraulic and transport parameters). Groundwater
age must, therefore, be regarded as a statistical, or probabilistic
distribution, rather than considering it as a single absolute or
average value. The dating methods commonly provide an average value
over the water sample for the age of groundwater after recharge,
which in theory does not represent hard data. In fact, the mean of
an unknown distribution, here the distribution of ages, is not
necessarily a reliable value for the most likely of this
distribution. By far the most frequently used dating methods are
based on the measurements of natural tracers, such as the isotopes
of radon, carbon or oxygen, and on the measurements of man-induced
atmospheric concentrations of elements such as tritium
($\nuc{3}{H}$), helium ($\nuc{3}{He}$), chloride ($\nuc{36}{Cl}$),
krypton ($\nuc{85}{Kr}$), or chlorofluorocarbons (CFC), which have
increased steadily between the 1940s and the early 1990s. The most
efficient methods for dating recent waters are the ones based on
$\nuc{3}{H}/\nuc{3}{He}$, $\nuc{85}{Kr}$, and CFC-measurements
\cite{Solomon92,Plummer93,Dunkle93}, which are known to provide age
dates over a period of ~40~years with an accuracy of 20\% or
less~\cite{Cook97}. CFC-based ages provide a good resolution for
groundwater with relatively young
components~\cite{Busenberg92,Boehlke96}, but they do not account for
the time spent by water molecules within the unsaturated zone, such
that in the case of deep water table aquifers the travel time
duration within the unsaturated zone must, therefore, be estimated.
Absolute dating techniques involve decay of radionuclides in
groundwater, for which $\nuc{14}{C}$ or $\nuc{36}{Cl}$ are classical
elements that are used for dating old groundwater, e.g. in deep and
large sedimentary basins. Groundwater age is very often used to make
inferences on aquifer parameters, groundwater recharge, flow paths
and flow velocities. However, with many of these dating methods, the
interpretation of ages is achieved by means of simple conceptual
models and fitting the data. This involves significant
simplifications of the flow and transport processes, which may lead
to erroneous interpretations about e.g. the past release of
contaminants in aquifers. As will be discussed in this paper, mixing
and dispersion are major factors, which can lead to unrepresentative
mean age measurements. Dating methods remain however very useful for
calibrating numerical models~\cite{Reilly94,Robertson89}, which
attempt to simulate the flow patterns, the flow rates, and the
distribution of ages in groundwater.

To quantify the distribution of ages in aquifers, several types of
mathematical models have been developed during the past decades of
research in this field. Nevertheless, in many cases age
distributions are more or less arbitrarily chosen and not
deterministically calculated. Analytical lumped-parameter type
models have been however extensively used in the simulation of
environmental tracer data
\cite{Maloszewski82,Maloszewski93,Zuber86,Campana84,Richter93,Amin96},
such as isotopic data, which are commonly interpreted with advective
transit time models, although isotope transport does not necessarily
undergo advective processes only. Specified transit time
distributions describing piston-flow, exponential mixing, combined
piston-exponential mixing, or dispersive mixing, are chosen to solve
the inverse problem by fitting the model on measured tracer output
data. This procedure calls for significant simplifications, which
can often not be justified, such as for instance neglecting the
reservoir structure, as well as the spatial variability of
infiltration rates~\cite{Campana87} and aquifer flow and transport
parameters. Amin and Campana~\cite{Amin96} proposed to model the
groundwater age mixing process by means of a three-parameter gamma
function which accounts for various states of mixing ranging between
no mixing (piston flow model) and perfect mixing (exponential
model). Robust verifications of the applicability of lumped
parameter models can hardly be
found~\cite{Haitjema95,Luther98,Etcheverry99}. Transit time
distributions are often obtained with numerical solutions by making
the assumption of pure advective motion of the groundwater
particles, the particle-tracking technique being the most popular
one. Purely kinematic ages ignore the effect of dispersion and
mixing on age transport~\cite{Davis82,Cordes92}, and often reveal to
be ill posed in complex heterogeneous systems~\cite{Varni98}, for
which the 3-D implementation is subject to severe technical
problems. The importance of including age dilution processes such as
dispersion and matrix diffusion, when comparison is made between
modeled and measured ages, has been pointed out by many
authors~\cite{Sudicky81,Maloszewski82,Maloszewski91}. Moreover,
particle-tracking does not allow calculation of transit time
distributions since groundwater volumes are not associated to
simulated ages.

More elaborated quantitative approaches consider age as a mass that
is transported by groundwater through volume-averaged temporal
moment equations \cite{Spalding58,Harvey95,Goode96,Varni98}, in
which the product of groundwater age with its mass (age mass $\rho
\textit{A}$) is the conserved quantity. Harvey and
Gorelick~\cite{Harvey95}, and Varni and Carrera~\cite{Varni98}
derived a set of recursive temporal moment equations, which are
sequentially solved in order to simulate the transit time
distributions from the $n$ calculated moments. According to Harvey
and Gorelick~\cite{Harvey95}, the first five moments that
characterize the accumulated mass, the mean, the variance, the
skewness and the kurtosis of a breakthrough curve, respectively, may
provide sufficient information to summarize the entire distribution.
Since many natural systems reveal a multi-modality of the age
distribution within the reservoir and at the discharge zones, and
since the shape of this distribution is a priori unknown, an
infinity of moments would therefore theoretically be required to
construct the entire distribution.

In the literature one can find many terms, which relate to a
specific time spent by water molecules within the aquifer. Use will
be made throughout this work of the notion of transit time as the
total residence time of water molecules within the aquifer, i.e. the
age of these molecules when they exit the aquifer. The notion of
travel time is rather used to characterize the time spent to travel
between two arbitrary locations in the aquifer. Travel time
probabilities have been a subject of high interest in many studies
characterizing solute transport in sub-surface
hydrology~\cite{Dagan82,Dagan87,Dagan89a,Jury82,Jury90}. The travel
time probability is commonly defined as the response function to an
instantaneous unit flux impulse~\cite{Danckwerts53,Jury90}. In their
transfer function approach of contaminant transport through
unsaturated soil units, Jury and Roth~\cite{Jury90} model tracer
breakthrough curves with one-dimensional travel time probability
functions. Shapiro and Cvetkovic~\cite{Shapiro88},
Dagan~\cite{Dagan89a}, and Dagan and Nguyen~\cite{Dagan89b} derived
the forward travel time probability for a mass of solute by using
the Lagrangian concept of particle displacement in porous media. The
derivation of forward and backward models for location and travel
time probability has become a classical mathematical approach for
contaminant transport characterization and
prediction~\cite{Uffink89,LaBolle98,Neupauer99,Neupauer01}. The
spreading of a contaminant mass is analyzed by following the random
motion of solute particles, and to do so, the advection--dispersion
equation (ADE) is assimilated to the Fokker-Planck (or forward
Kolmogorov) equation. The expected resident concentration of a
conservative tracer is taken as the probability density function for
the location of a particle, at any time after having entered the
system.

The aim of the present work is to provide a general theoretical
framework to model complete groundwater age distributions at aquifer
scale in a deterministic way. The concept of age variability is
associated to the concept of random variables and their
distributions by using classical elements of probability, allowing
the introduction of mathematical definitions for age, life
expectancy and transit time statistical distributions. Forward and
backward ADEs for conservative tracers are used to simulate the
above-mentioned distributions at aquifer scale. By manipulating the
ADEs, the reservoir theory~\cite{Eriksson71} is expressed in order
to characterize recharge and discharge zones transit time
distributions with refined accuracy. The proposed models are
illustrated and discussed by means of analytical and numerical
analysis of one- and two-dimensional theoretical flow
configurations.
\section{The 'ages' of groundwater as space-dependent random variables.}
In this section we present the models allowing the calculation of
the statistical distribution of groundwater age, life expectancy and
transit time in arbitrary aquifers.
\subsection{Definitions.}
The characterization of groundwater molecules with respect to a
travel time within an aquifer system is fully dependent on the
spatial reference from which this time is "measured". Usually the
groundwater age is defined as a relative quantity with respect to a
starting location where age is assumed to be zero. Use will be made
throughout this work of three variations of terminology. For a given
spatial position in the reservoir, the age (\textit{A}) relates to
the time elapsed since the water molecules entered the system at the
recharge limits, where age is zero. For the same spatial position,
the life expectancy (\textit{E}) is defined as the time required for
the water molecules to reach an outlet limit of the system. Life
expectancy is therefore zero at an outlet. The transit time
(\textit{T}) finally refers to the total time required by the same
water molecules to migrate from an inlet zone ($T = E$) to an outlet
zone ($T = A$). In a REV, the three variables \textit{A}, \textit{E}
and \textit{T} are random variables, characterized by probability
density functions (pdf) that can be regarded as the statistical
occurrence of water molecules with respect to time, which could be
observed in a groundwater sample if any analytical procedure would
allow such measurements.
\subsection{Age probability.}
The typical heterogeneity of aquifer systems involves strongly
varying flow velocity fields, with multi-scale coherence lengths.
The spatial variability of velocity and transport parameters induces
a spreading of the contaminant mass. The tensor of macro-dispersion
in the classical ADE accounts for the uncertainty in the transport
prediction induced by mixing. Various studies relate to how the ADE
is limited by the impact of physical and chemical heterogeneities on
solute transport, such that up-scaling is not always
satisfying~\cite{Dagan84,Sudicky86,Gelhar83}. If such
heterogeneities are present at aquifer scale the transport
parameters should be time-dependent, but this time dependency may be
relaxed when the correlation scales of the transport parameter
random fields are finite~\cite{Gelhar83,Neumann87}. The ADE with
time-independent parameters holds only when the solute particles
have enough time to be distributed by dispersion between the flow
lines. Since we are interested in solving the age transport problem
at aquifer scale, we make the assumption that the ADE with
time-independent transport parameters (the parameters have reached
their asymptotic values) can model the evolutional transport of the
groundwater age and life expectancy distributions under steady-flow
conditions. The modelled process applies to conservative and
non-reactive tracers.

\begin{figure}[tbp]
\begin{center}
\includegraphics[width=\textwidth]{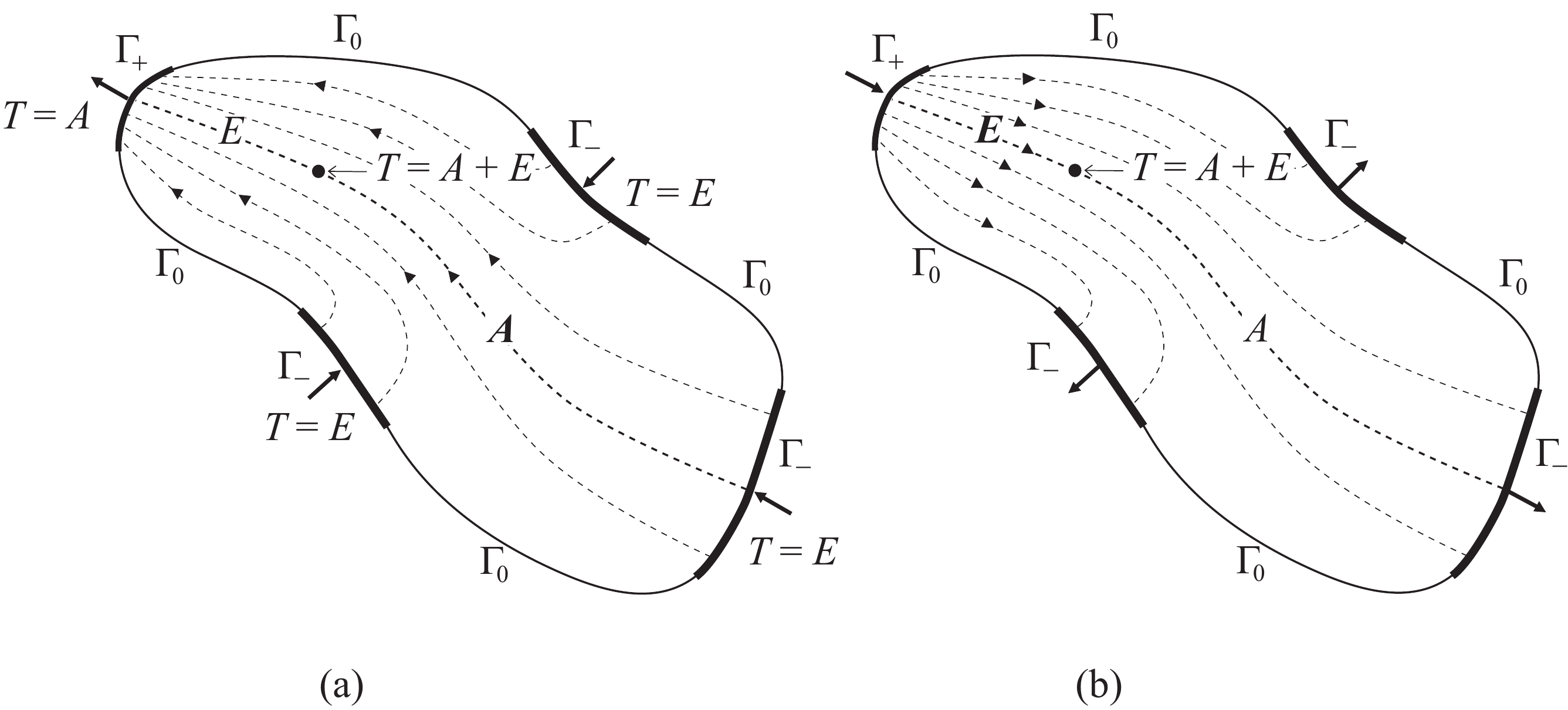}
\end{center}
\caption{Schematic illustration of a groundwater reservoir $\Omega$,
with indicated no-flow ($\Gamma_0$), inlet ($\Gamma_-$) and outlet
($\Gamma_+$) boundaries: (a) Age problem with normal flow field; (b)
Life expectancy problem with reversed flow field. The arrow heads on
the symbolized flowlines (dashed lines) stand for the position and
direction of water molecules at a given time after their release.
The black dot stands for a small water sample, to illustrate the
random variable transit time (\textit{T}) as the sum of the two
random variables age (\textit{A}) and life expectancy
(\textit{E}).\label{fig:F1}}
\end{figure}

Let us consider an aquifer domain $\Omega$ in the three-dimensional
space, with hydraulic recharge boundary $\Gamma_-$, discharge
boundary $\Gamma_+$, and impermeable boundary $\Gamma_0$, as
illustrated in Fig.~\ref{fig:F1}. The boundary $\Gamma_+$
corresponds to the open boundary of the system, through which a free
exit of age mass is expected. With respect to the above-mentioned
considerations on contaminant spreading, it is convenient to
describe the groundwater sample age distribution as a random
variable associated to a probabilistic model. The age probability
distribution at a position \textit{\textbf{x}} in $\Omega$ can be
evaluated by solving the ADE when a unit pulse of conservative
tracer is uniformly applied on the recharge area $\Gamma_-$. The
resulting breakthrough curve is the probabilistic age
distribution~\cite{Danckwerts53,Jury90}. Making use of this
property, we propose to model groundwater age and life expectancy
pdfs by forward and backward transient-state transport equations,
under steady-state hydraulic conditions. The pre-solution of the
velocity field is performed by the following steady-state
groundwater flow equation:
\begin{equation}\label{eq:flow}
\nabla\cdot\textbf{q} = q_\textrm{I} - q_\textrm{O}, \quad
\textbf{q} = -\textbf{K}\nabla\textit{H} \quad\quad \text{ in }
\quad \Omega,
\end{equation}
where \textbf{q} is the Darcy flux vector
$[\textrm{L}\textrm{T}^{-1}]$, which is valid for ideal tracers,
\textit{H} is the hydraulic head $[\textrm{L}]$, $q_\textrm{I}$ and
$q_\textrm{O}$ are fluid source and sink terms $[\textrm{T}^{-1}]$,
respectively, and \textbf{K} is the tensor of hydraulic conductivity
$[\textrm{L}\textrm{T}^{-1}]$. The age pdf is then obtained by
solving the following forward boundary value problem:
\begin{subequations}\label{eq:A BVP}
\begin{equation}\label{eq:A ADE}
\frac{\partial \phi g_A}{\partial t} = -\nabla\cdot\textbf{q} g_A +
\nabla\cdot \textbf{D}\nabla g_A + q_\textrm{I} \delta(t) -
q_\textrm{O} g_A \quad\quad \text{ in } \quad \Omega
\end{equation}
\begin{equation}\label{eq:IC A ADE}
g_A(\textit{\textbf{x}},0) = g_A(\textit{\textbf{x}},\infty) = 0
\quad\quad \text{ in } \quad \Omega
\end{equation}
\begin{equation}\label{eq:BC1 A ADE}
\textbf{J}_A(\textit{\textbf{x}},t)\cdot \textit{\textbf{n}} =
(\textbf{q}\cdot \textit{\textbf{n}})\delta(t) \quad\quad \text{ on
} \quad \Gamma_-
\end{equation}
\begin{equation}\label{eq:BC2 A ADE}
\textbf{J}_A(\textit{\textbf{x}},t)\cdot \textit{\textbf{n}} = 0
\quad\quad \text{ on } \quad \Gamma_0
\end{equation}
\end{subequations}

where $g_A(\textit{\textbf{x}},t)$ is the transported age pdf
$[\textrm{T}^{-1}]$, $\textbf{J}_A(\textit{\textbf{x}},t)$ is the
total age mass flux vector $[\textrm{L}\textrm{T}^{-2}]$, \textbf{D}
is the tensor of macro-dispersion $[\textrm{L}^{2}\textrm{T}^{-1}]$,
$\textit{\textbf{x}} = (\textit{x},\textit{y},\textit{z})$ is the
vector of cartesian coordinates $[\textrm{L}]$, \textit{t} is time
$[\textrm{T}]$, $\phi = \phi(\textit{\textbf{x}})$ is porosity or
mobile water content $[-]$, \textit{\textbf{n}} is a normal outward
unit vector, and $\delta(t)$ is the time-Dirac delta function
$[\textrm{T}^{-1}]$, which ensures a pure flux impulse on
$\Gamma_-$. The source term $q_\textrm{I} \delta(t)$ is meant for
simulating a potential internal production of water (3-D) or 2-D
horizontal aquifer configurations with an areal recharge intensity
$q_\textrm{I}$. The sink term $q_\textrm{O} g_A$ may result from any
internal extraction of groundwater. The tensor of macro-dispersion
$\textbf{D} = \phi \textbf{D}^* = \textbf{D}(\textit{\textbf{x}})$
in Eq.~(\ref{eq:A ADE}) is defined by Bear~\cite{Bear72}:
\begin{equation}\label{eq:Dtensor}
\textbf{D} = (\alpha_L - \alpha_T) \frac{\textbf{q}\otimes
\textbf{q}}{\|\textbf{q}\|} + \alpha_T \|\textbf{q}\|\textbf{I} +
\phi D_\textrm{m} \textbf{I}
\end{equation}

where $\alpha_L$ and $\alpha_T$ are the longitudinal and transversal
coefficients of dispersivity $[\textrm{L}]$, respectively,
$D_\textrm{m}$ is the coefficient of molecular diffusion
$[\textrm{L}^{2}\textrm{T}^{-1}]$, and $\textbf{I}$ is the identity
matrix. The total age mass flux vector
$\textbf{J}_A(\textit{\textbf{x}},t)$ is classically defined by the
sum of the convective and dispersive fluxes:
\begin{equation}\label{eq:age flux}
\textbf{J}_A(\textit{\textbf{x}},t) = \textbf{q}
g_A(\textit{\textbf{x}},t) -\textbf{D}\nabla
g_A(\textit{\textbf{x}},t)
\end{equation}
\subsection{Life expectancy probability.}
The life expectancy probability distribution satisfies the adjoint
backward model of Eq.~(\ref{eq:A ADE}):
\begin{subequations}\label{eq:LE BVP}
\begin{equation}\label{eq:LE ADE}
\frac{\partial \phi g_E}{\partial t} = \nabla\cdot\textbf{q} g_E +
\nabla\cdot \textbf{D}\nabla g_E - q_\textrm{I} g_E \quad\quad
\text{ in } \quad \Omega
\end{equation}
\begin{equation}\label{eq:IC LE ADE}
g_E(\textit{\textbf{x}},0) = g_E(\textit{\textbf{x}},\infty) = 0
\quad\quad \text{ in } \quad \Omega
\end{equation}
\begin{equation}\label{eq:BC1 LE ADE}
\textbf{J}_E(\textit{\textbf{x}},t)\cdot \textit{\textbf{n}} =
-(\textbf{q}\cdot \textit{\textbf{n}})\delta(t) \quad\quad \text{ on
} \quad \Gamma_+
\end{equation}
\begin{equation}\label{eq:BC2 LE ADE}
-\textbf{D}\nabla g_E(\textit{\textbf{x}},t)\cdot
\textit{\textbf{n}} = 0 \quad\quad \text{ on } \quad \Gamma_0
\end{equation}
\end{subequations}

where $g_E(\textit{\textbf{x}},t)$ is the transported life
expectancy pdf, and where the total life expectancy mass flux vector
$\textbf{J}_E(\textit{\textbf{x}},t)$ is
\begin{equation}\label{eq:LE flux}
\textbf{J}_E(\textit{\textbf{x}},t) = -\textbf{q}
g_E(\textit{\textbf{x}},t) -\textbf{D}\nabla
g_E(\textit{\textbf{x}},t)
\end{equation}

Eq.~(\ref{eq:LE ADE}) is the formal adjoint of Eq.~(\ref{eq:A
ADE})~\cite{Garabedian64,Arnold74}, the so-called "backward-in-time"
equation~\cite{Uffink89,Wilson97,Weissmann02}, also backward
Kolmogorov equation~\cite{Kolmogorov31}. Given the forward equation,
the backward equation is technically obtained by reversing the sign
of the flow field, and by adapting the boundary
conditions~\cite{Neupauer99,Neupauer01,Weissmann02}. On the
impermeable boundary $\Gamma_0$, a third-kind condition (Cauchy) in
the forward equation becomes a second-kind condition (Neumann) in
the backward equation, and vice-versa. A second-kind condition in
the forward model will also become a third-kind condition in the
backward model~\cite[p.~146]{Gardiner83}. The advection term is
known to be not self-adjoint (it should be written in the form
$\textbf{q}\cdot\nabla g_E$ in Eq.~(\ref{eq:LE ADE})) unless flow is
divergence free~\cite{Weissmann02}. However, the backward equation
can still handle non-divergence free flow fields by means of the
important sink term $- q_\textrm{I} g_E$ appearing in
Eq.~(\ref{eq:LE ADE}). This sink term has been derived in
Cornaton~\cite{Cornaton03} from the vertical averaging process of
the general 3-D backward ADE, and is consistent with the analysis of
Neupauer and Wilson~\cite{Neupauer01,Neupauer02}. Recharge by
internal sources (3-D or 2-D vertical) or by areal fluxes (fluid
source for 2-D horizontal) is introduced by the first-order decay
type term $- q_\textrm{I} g_E$, which is a consequence of the
reversed flow field. Internal sources produce a sink of life
expectancy probability, while internal sinks (term $q_\textrm{O}
g_A$ in Eq.~(\ref{eq:A ADE})) do not appear in the backward model
since a fluid sink may not influence the life expectancy pdf. The
boundary $\Gamma_-$ corresponds to the open boundary of the system
(since flow is reversed) through which a free exit of life
expectancy mass is expected. \smallskip \\
The simulation of life expectancy with~(\ref{eq:LE BVP}) is valid in
the case of steady-state velocity fields only. Transient-state
velocity fields would require another appropriate formulation
of~(\ref{eq:LE BVP}). For a steady-state hydraulic situation, if
\textbf{q} approaches zero in some regions of the reservoir, e.g.
like in aquitards in which transport is diffusion-dominant,
then~(\ref{eq:LE BVP}) still holds because of the irreversible
nature of dispersion. The amount of age mass diffused between an
aquifer and an aquitard is proportional to concentration differences
between the two media, and is the same in both the forward and
backward problems.
In the boundary value problems~(\ref{eq:A BVP}) and~(\ref{eq:LE
BVP}) the classical homogeneous Neumann boundary condition ($-
\textbf{D}\nabla g \cdot \textit{\textbf{n}} = 0$) at the outlet
limit for the age problem (at the inlet limit for the life
expectancy problem) is not used in order to allow a natural age/life
expectancy gradient through the open boundaries. Instead, the normal
projection of the dispersive flux is evaluated implicitly at the
boundaries. The evaluation procedure in the framework of the finite
element method is described in Cornaton et al.~\cite{Cornaton04}.
This kind of boundary condition, which is referred to as
\emph{Implicit Neumann condition}, enables continuity of the total
mass flux at outlet. The \emph{Implicit Neumann condition} is a
generalized version of the \emph{Free Exit condition} for parabolic
equations proposed by Frind~\cite{Frind88}. As discussed by Nauman
and Buffham ~\cite{Nauman83}, Parker and van
Genuchten~\cite{Parker84}, Kreft and Zuber~\cite{Kreft86}, and Bear
and Verruitj~\cite{Bear87}, total mass flux continuity at outlet
permits upgradient solute movement by dispersion.

Eqs.~(\ref{eq:A BVP}) and (\ref{eq:LE BVP}) simulate the forward and
backward transport resulting from a unit pulse input. The space-time
evolution of the water molecules is described by the distributions
$g_A(\textit{\textbf{x}},t)$ and $g_E(\textit{\textbf{x}},t)$. Both
differential equations deal with conditional probabilities that
characterize the statistical occurrence of water molecules as a
function of age and life expectancy. Location probability is related
to resident concentration~\cite{Kreft78,Dagan87,Jury90}, and
describes the position \textit{\textbf{x}} of water molecules at a
given time after their release in the system. On the other hand,
travel time probability is related to flux
concentration~\cite{Kreft78,Dagan87,Jury90}, and characterizes for a
position \textit{\textbf{x}} the amount of time spent within
$\Omega$ since the water molecules entered the system (in the
right-hand side of Eq.~(\ref{eq:BC1 LE ADE}), age is at the flux
concentration $g^f_A = \delta(t)$). Resident concentration relates
to the mass of solute per unit porous volume while flux
concentration is defined as the solute mass flux per unit water
flux. A flux concentration is the physical representation of the
mean of the microscopic fluid concentrations weighted by the
associated microscopic fluid velocities~\cite{Parker84}. The
multi-dimensional relation between flux and resident concentrations
can be found in Sposito and Barry~\cite{Sposito87}, and is formally
the projection of the total mass flux on the flow velocity
direction. Accordingly, the flux concentration form of the random
variable age is:
\begin{equation}\label{eq:A flux conc}
g_A^f =
\frac{\textbf{J}_A\cdot\textbf{q}}{\parallel\textbf{q}\parallel^2} =
g_A +
\frac{\textbf{J}_A^d\cdot\textbf{q}}{\parallel\textbf{q}\parallel^2}
= g_A - \frac{\textbf{D}\nabla
g_A\cdot\textbf{q}}{\parallel\textbf{q}\parallel^2}
\end{equation}

with $\textbf{J}^d =-\textbf{D}\nabla g$ denoting the dispersive
part of \textbf{J}. By analogy, the flux concentration form of the
variable \textit{E} can be defined as
\begin{equation}\label{eq:LE flux conc}
g_E^f = -
\frac{\textbf{J}_E\cdot\textbf{q}}{\parallel\textbf{q}\parallel^2} =
g_E -
\frac{\textbf{J}_E^d\cdot\textbf{q}}{\parallel\textbf{q}\parallel^2}
= g_E + \frac{\textbf{D}\nabla
g_E\cdot\textbf{q}}{\parallel\textbf{q}\parallel^2}
\end{equation}

The age pdf at a position \textbf{\textit{x}} in $\Omega$
characterizes the probability per unit time for the time spent since
recharge, and is a flux concentration, evaluated enforcing
Eq.~(\ref{eq:A flux conc}). The resident concentration of age
characterizes a volume-averaged age density function, which
describes the density of probability of finding water molecules at
the position \textit{\textbf{x}} in $\Omega$, at time \textit{t}.
\subsection{Transit time probability.}
In Eqs.~(\ref{eq:A ADE}) and (\ref{eq:LE ADE}), the dependent
variables are probability density functions of continuous time
random variables. The behavior of these random variables can also be
described by cumulative distribution functions (cdf). Let \textit{U}
be one of these two variables ($\textit{U} = \textit{A}$ or
$\textit{U} = \textit{E}~$), with \textit{u} the associated values
the variable \textit{U} may take at a given position
\textit{\textbf{x}} of space. The cdf $G_U(\textit{\textbf{x}},u)$
and the pdf $g_U(\textit{\textbf{x}},u)$ of the variable \textit{U}
are commonly defined as
\begin{equation}\label{eq:CDFdef}
G_U (\textit{\textbf{x}},u) = {\rm{P}}\left[ { - \infty \le U \le u}
\right] = \int_{ - \infty }^u {g_U (\textit{\textbf{x}}},\tau )d\tau
\end{equation}
with
\begin{equation}\label{eq:PDFdef}
g_U (\textit{\textbf{x}},u) = \frac{{\partial G_U
(\textit{\textbf{x}},u)}} {{\partial u}} \quad \text{, } \quad G_U
(\textit{\textbf{x}},- \infty ) = 0 \quad \text{, } \quad G_U
(\textit{\textbf{x}},\infty ) = 1
\end{equation}

where P denotes the probability event on \textit{U}, or number of
occurrences with $\textit{U} \leq u$ ratioed to the total number of
occurrences, and $\tau$ is a dummy variable for integration. The
probability functions property~(\ref{eq:PDFdef}) together with the
boundary conditions in Eqs.~(\ref{eq:A BVP}) and (\ref{eq:LE BVP})
ensure that age and life expectancy $g_A$ and $g_E$ are directly
related to probabilities, and since concentration can be modelled by
the ADE, then probability too can be modelled by the ADE. For a
given position \textit{\textbf{x}} in $\Omega$, the ages of
groundwater molecules are described by the pdf $g_A$, which measures
the density of probability of having an age \textit{t}. The same
molecules are also characterized by the pdf $g_E$, which measures
the density of probability of having a life expectancy \textit{t}.
Introducing now the random variable transit time \textit{T}, with
density of probability $g_T$, the water molecules can be described
by their intensity of probability of flowing throughout the system
at a time \textit{t}. The variable \textit{T} is a random variable
corresponding to the sum of the two random variables \textit{A} and
\textit{E} (see Fig.~\ref{fig:F1}). Hence the statistical
distribution of \textit{T} is the pdf of the sum of \textit{A} and
\textit{E}, $g_T = g_{A + E}$. This problem can be solved if the
joint pdf $g_{A,E}$ of \textit{A} and \textit{E}, which
characterizes the joint behavior of \textit{A} and \textit{E}, is
known~\cite{Benjamin70}:
\begin{equation}\label{eq:jointPDFae}
g_T (\textit{\textbf{x}},t) = g_{A + E} (\textit{\textbf{x}},t) =
\int_{ - \infty }^{ + \infty } {g_{A,E} (\textit{\textbf{x}},\tau ,t
- \tau ) d\tau }
\end{equation}

The joint quantity $g_{A,E}(\textit{\textbf{x}},a,e)dade$ relates to
the probability that \textit{A} lies in the small interval
\textit{a} to $a + da$, and that \textit{E} lies in the small
interval \textit{e} to $e + de$. Assuming that \textit{A} and
\textit{E} are stochastically independent variables (for the same
spatial position, \textit{A} depends on the initial point while
\textit{E} depends on the end point, and under a steady-state flow
regime \textit{E} may not be influenced by the memory of the past
evolution), and since $g_A$ and $g_E$ are zero for negative values
of their arguments, the joint pdf $g_{A,E}$ simplifies in
$g_Ag_E$~\cite{Benjamin70}. The probability density function $g_T$
in Eq.~(\ref{eq:jointPDFae}) can then be obtained using the
convolution integral:
\begin{equation}\label{eq:TT PDF}
g_T (\textit{\textbf{x}},t) = \int_0^t {g_A
(\textit{\textbf{x}},\tau )g_E (\textit{\textbf{x}},t - \tau ) d\tau
}
\end{equation}

from which the cdf of \textit{T} can be calculated enforcing
Eq.~(\ref{eq:CDFdef}). The fact that $g_A$ and $g_E$ are zero for
negative values of their arguments allows applying the convolution
integral from 0 to \textit{t}. Since the pdfs $g_A$ and $g_E$ give
the age and life expectancy probability of occurrence at each
position \textit{\textbf{x}} in $\Omega$, both the maximum age as
well as the maximum life expectancy correspond to the maximum
transit time. Consequently, the time variable \textit{t} can
equivalently refer to all specific values of age, life expectancy,
and transit time. The convolution integral~(\ref{eq:TT PDF}) states
that the probability that the variable \textit{T} lies in a small
interval around \textit{t} is proportional to the product of the
probability that the variable \textit{A} lies in the interval
\textit{t} to $t + dt$ and a factor proportional to the probability
that the variable \textit{E} lies in a small interval around $t -
\tau$, the value of \textit{E} ensuring that \textit{A} and
\textit{E} sum to \textit{T}. This product is then summed over all
possible values of time \textit{t} (from the minimum age to the
maximum age) to yield the transit time pdf at a position
\textit{\textbf{x}} in space. The derived distribution of $T = A +
E$ in Eq.~(\ref{eq:TT PDF}) can also be viewed as a transfer
function convolution process, the input distribution being the age
pdf $g_A$, and the signal transferring function being the life
expectancy $g_E$.

To our point of view, Eq.~(\ref{eq:TT PDF}) is an important result
of the present work. The field of $g_T$ characterizes the evolution
of groundwater molecules throughout the aquifer domain by specifying
the amount of time from recharge to discharge. At a given position
in the reservoir, the temporal evolution of the groundwater
molecules can be characterized by the three pdfs $g_A$, $g_E$ and
$g_T$. Each function contains specific information on a time of
residence, the nature of which is a function of the spatial
references that are chosen for evaluation. For instance, $g_A$ is
conditioned by the inlet limit $\Gamma_-$, where the variable
\textit{A} is nil, while $g_E$ is conditioned by the outlet limit
$\Gamma_+$, where the variable \textit{E} is nil. For the variable
\textit{T}, the pdf $g_T$ is conditioned by the fact that $T = A$ at
outlet, and that $T = E$ at inlet.

If $g_A$ and $g_E$ are resident concentrations, so is $g_T$. If
$g_A$ and $g_E$ are flux concentrations, so is $g_T$. Applying a
Laplace transform to Eqs.~(\ref{eq:A flux conc}) and (\ref{eq:LE
flux conc}) and convoluting, the transit time resident and flux
concentrations are found to be linked by the following relation:
\begin{equation}\label{eq:TT flux conc}
\hat g_T^f = \hat g_A^f \hat g_E^f = \hat g_T - \frac{1}{{\left\|
{\bf{q}} \right\|^2 }}[\hat g_A {\bf{\hat J}}_E^d - \hat g_E
{\bf{\hat J}}_A^d + \frac{{({\bf{\hat J}}_A^d \otimes {\bf{\hat
J}}_E^d ){\bf{q}}}}{{\left\| {\bf{q}} \right\|^2 }}] \cdot {\bf{q}}
\end{equation}

where $\hat g_U (\textit{\textbf{x}},s)$ denotes the
\textit{s}-transform state of the function
$g_U(\textit{\textbf{x}},u)$, \textit{U} = \textit{A}, \textit{E} or
\textit{T}. Eq.~(\ref{eq:TT flux conc}) shows that transit time flux
concentration is dependent on the transit time resident
concentration, but also on the tensor product between the age and
life expectancy dispersive fluxes, and on the convolution product of
the age and life expectancy flux and resident concentrations. For a
zero dispersion case, $g^f = g$ for the random variables \textit{A},
\textit{E} and \textit{T}.
\begin{figure}[tbp]
\begin{center}
\includegraphics[width=\textwidth]{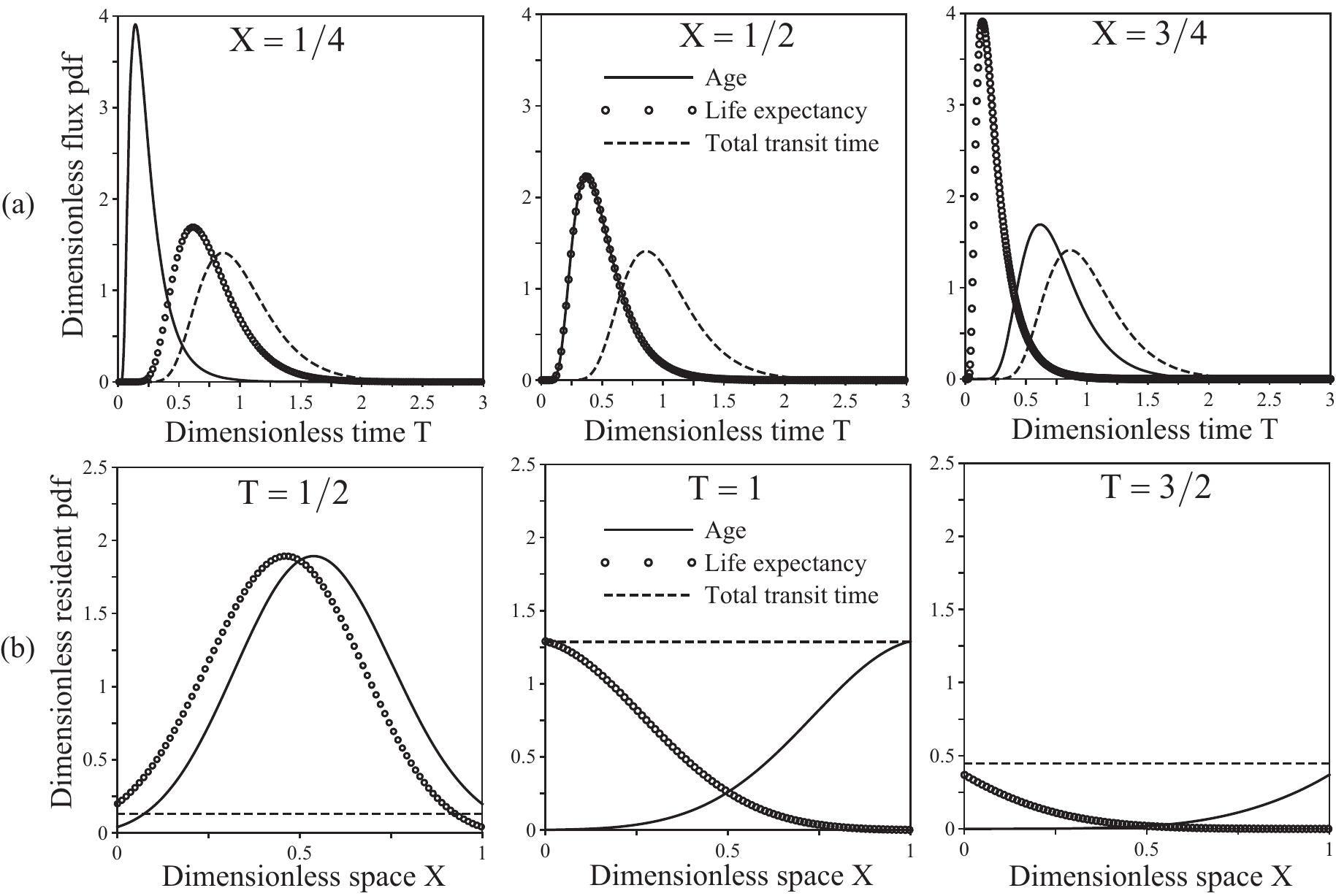}
\end{center}
\caption{Age, life expectancy and transit time dimensionless pdfs in
a 1-D domain for a P\'{e}clet number Pe = 20: (a) Flux pdfs; (b)
Resident pdfs. Time is normalized by the average turnover time
$\tau_0 = L/v$, \textit{x} by \textit{L}, and $\textrm{Pe} = Lv/D$.
The average age and the average life expectancy sum to the average
transit time.\label{fig:F2}}
\end{figure}
Consider the semi-infinite 1-D domain of characteristic length
\textit{L} (the outlet is supposed at the position $x = L$) and
uniform velocity \textit{v} along the \textit{x}-axis, as
illustrated in Fig.~\ref{fig:F2}. The age flux pdf at the position
\textit{x} is obtained as the solution of Eq.~(\ref{eq:A ADE}) using
the boundary condition $g_A = g^f_A = \delta(t)$ at $x = 0$, and the
age resident pdf is obtained as the solution of Eq.~(\ref{eq:A ADE})
using the boundary condition~(\ref{eq:BC1 A ADE}) at $x = 0$. These
solutions are given in dimensionless form in Appendix~A. They show
the trivial fact that for a one-dimensional flow configuration, the
transit time flux pdf is unique and independent on the spatial
coordinate, as illustrated in Fig.~\ref{fig:F2}a. For example, the
age flux pdf at $\textrm{X} = 1/4$ in Fig.~\ref{fig:F2}a is equal to
the life expectancy flux pdf at $\textrm{X} = 3/4$, and the
convolution of both distributions gives the transit time flux pdf,
which equals both the age flux pdf at outlet ($\textrm{X} = 1$) and
the life expectancy flux pdf at inlet ($\textrm{X} = 0$). The
average age $x/v$ and the average life expectancy $(L - x)/v$ sum to
the average transit time $L/v$, which is independent on \textit{x}
in a rectilinear flow line. In a similar way, the age resident pdf
at the position $\textrm{X}$ equals the life expectancy resident pdf
at the position $1 - \textrm{X}$, as shown in Fig.~\ref{fig:F2}b.
For a fixed value of time, the transit time resident pdf is
constant. This shows that, in 1-D, the intensity of probability of
finding water molecules at any position in the domain that transit
at time \textit{t} or less, given that \textit{t} is fixed, is
always identical. The transit time resident pdf gives the intensity
of probability for the spatial position of water molecules, for a
given value of transit time. In 1-D, this intensity of probability
is uniform since the trajectory is unique. The resident age and life
expectancy curves show the typical apparent discontinuity in
concentration at inlet (resident pdf of \textit{A}) and at outlet
(resident pdf of \textit{E}). These discontinuities have a drawback
on the transit time resident pdf, which is not necessarily equal to
the age resident pdf at outlet, and equivalently not necessarily
equal to the life expectancy resident pdf at inlet. This can be
attributed to dispersion effects at the boundaries. Since the Cauchy
condition is homogeneous for $\textrm{T} = 0^{+}$ at inlet for the
age pdf problem, and at outlet for the life expectancy pdf problem,
backward movement of water molecules by dispersion (i.e. in the
reserved direction of velocity) is put down by non-zero age and life
expectancy resident concentrations at the boundaries, the magnitude
of which is higher the higher the dispersivity. If we consider e.g.
the inlet boundary, the age and life expectancy concentrations are
both not nil. The convolution of both pdfs is, therefore, not equal
to any of them (age pdf at outlet, life expectancy pdf at inlet)
since both concentrations can have a significant value at inlet and
outlet at the same time.

The spatial distribution of the transit time pdf is also ruled by a
differential equation. Combining Eqs.~(\ref{eq:A ADE}) and
(\ref{eq:LE ADE}) after a Laplace transform, the following equation
can also be obtained:
\begin{subequations}
\begin{equation}\label{eq:TT ADE}
\nabla \cdot \textbf{q} g_T = \mathcal{L}^{- 1} \left\{ \hat S_d
\right\}
\end{equation}
\begin{equation}\label{eq:TT ADE source}
\hat S_d = \hat g_E \nabla \cdot {\bf{\hat J}}_A^d - \hat g_A \nabla
\cdot {\bf{\hat J}}_E^d
\end{equation}
\end{subequations}

where $\mathcal{L}^{- 1}$ denotes the inverse Laplace transform.
Eq.~(\ref{eq:TT ADE}) is the transit time pdf differential equation.
It is of steady-state kind, with a source term that accounts for the
divergence of the age and life expectancy dispersive fluxes. For the
pure advective case in divergence-free flow fields, Eq.~(\ref{eq:TT
ADE}) simplifies in $\textbf{q}\cdot\nabla
g_T(\textit{\textbf{x}},t) = 0$. This local condition states that
the flux vector and the transit time gradient have to be always
perpendicular, as a requirement for keeping transit times constant
along the flow paths. The resolution of Eq.~(\ref{eq:TT ADE}), which
is of hyperbolic kind, is linked to technical difficulties, e.g. for
the evaluation of the source term~(\ref{eq:TT ADE source}) over the
domain, and is not beneficial since the pdfs
$g_A(\textit{\textbf{x}},t)$ and $g_E(\textit{\textbf{x}},t)$ must
be evaluated in a preliminary step.
\subsection{Mean age, mean life expectancy and mean transit time.}
The average values of the probability density functions $g_A$, $g_E$
and $g_T$ are defined by their first order temporal moment, the
$n^\textrm{th}$ moment being
\begin{equation}\label{eq:nmoment}
\mu_n {\rm{[}}g_U {\rm{]}} = \int_{ - \infty }^{ + \infty } {u^n g_U
(\textit{\textbf{x}},u)du} = ( - 1)^n \frac{{\partial ^n \hat g_U
(\textit{\textbf{x}},s = 0)}}{{\partial s^n }}
\end{equation}

with \textit{U} = \textit{A}, \textit{E}, or \textit{T}. Applying
the convolution theorem in the Laplace domain results in the
transformed Eq.~(\ref{eq:TT PDF}), $\hat g_T (\textit{\textbf{x}},s)
= \hat g_A (\textit{\textbf{x}},s) \hat g_E
(\textit{\textbf{x}},s)$. Accounting for the pdf property $\hat g_U
(\textit{\textbf{x}},s = 0) = 1$ and enforcing
Eq.~(\ref{eq:nmoment}) for \textit{n} = 1 and 2 yields the average
transit time and its variance:
\begin{equation}\label{eq:meanTT}
\langle T\rangle  = \langle A\rangle + \langle E\rangle
\end{equation}
\begin{equation}\label{eq:varTT}
\mu _{\rm{2}} {\rm{[}}g_T {\rm{]}} = \mu _{\rm{2}} {\rm{[}}g_A
{\rm{]}} + \mu _{\rm{2}} {\rm{[}}g_E {\rm{]}} {\rm{ + }} {\rm{2}}
\langle A\rangle \langle E\rangle
\end{equation}
\begin{equation}\label{eq:stdTT}
\sigma ^2 {\rm{[}}g_T {\rm{]}} = \sigma ^2 {\rm{[}}g_A {\rm{]}} +
\sigma ^2 {\rm{[}}g_E {\rm{]}}
\end{equation}

with $\langle A\rangle = \langle A\rangle(\textit{\textbf{x}}) =
\mu_1[g_A]$, $\langle E\rangle = \langle
E\rangle(\textit{\textbf{x}}) = \mu_1[g_E]$ and $\langle T\rangle =
\langle T\rangle(\textit{\textbf{x}}) = \mu_1[g_T]$. The average age
and average life expectancy of a water sample sum up to the average
transit time. Since the variables $A$ and $E$ are supposed to be
stochastically independent, the variances $\sigma^2$ of the age and
life expectancy pdfs also sum up to the variance of the transit time
pdf. Using the operator in Eq.~(\ref{eq:nmoment}), Eqs.~(\ref{eq:A
ADE}) and~(\ref{eq:LE ADE}) can be transformed into their
$n^\textrm{th}$ normalized moment $\mu_n$ form. The recursive
application of the operator ~(\ref{eq:nmoment}) to Eqs.~(\ref{eq:A
ADE}) and~(\ref{eq:LE ADE}) yields
\begin{equation}\label{eq:nmoment A ADE}
- \nabla  \cdot {\bf{q}} \mu_n [g_A ] + \nabla  \cdot {\bf{D}}\nabla
\mu_n [g_A ] - q_\textrm{O} \mu_n [g_A ] + \phi n \mu_{n - 1} [g_A ]
= 0
\end{equation}
for the forward $n^\textrm{th}$ moment ADE, and
\begin{equation}\label{eq:nmoment E ADE}
\nabla  \cdot {\bf{q}} \mu_n [g_E ] + \nabla  \cdot {\bf{D}}\nabla
\mu_n [g_E ] - q_\textrm{I} \mu_n [g_E ] + \phi n \mu_{n - 1} [g_E ]
= 0
\end{equation}

for the backward $n^\textrm{th}$ moment ADE. The $n^\textrm{th}$
moment forms of the ADEs~(\ref{eq:A ADE}) and~(\ref{eq:LE ADE}) are
only dependent on the $(n - 1)^\textrm{th}$ moment $\mu_{n-1}$. For
instance, since $\mu_0[g_A] = 1$ and $g_A(\textit{\textbf{x}},0) =
0$, the first moment form of Eq.~(\ref{eq:A ADE}) (for $n = 1$)
corresponds to the mean age equation as defined by
Goode~\cite{Goode96}, in which the mean age is the average over a
sample of water molecules of the time elapsed since recharge:
\begin{equation}\label{eq:mean A ADE}
- \nabla  \cdot {\bf{q}} \langle A\rangle + \nabla  \cdot
{\bf{D}}\nabla \langle A\rangle - q_\textrm{O} \langle A\rangle +
\phi = 0 \quad\quad \text{ in } \quad \Omega
\end{equation}

The first moment form of Eq.~(\ref{eq:LE ADE}) gives the backward
adjoint mean life expectancy equation:
\begin{equation}\label{eq:mean LE ADE}
\nabla  \cdot {\bf{q}} \langle E\rangle + \nabla  \cdot
{\textbf{D}}\nabla \langle E\rangle - q_{\textrm{I}} \langle
E\rangle + \phi  = 0 \quad\quad \text{ in } \quad \Omega
\end{equation}

Finally, the mean transit time equation is deduced by subtracting
Eq.~(\ref{eq:mean A ADE}) and Eq.~(\ref{eq:mean LE ADE}):
\begin{subequations}
\begin{equation}\label{eq:mean TT ADE}
{\bf{q}} \cdot \nabla \langle T\rangle  = \langle S_d \rangle
\end{equation}
\begin{equation}\label{eq:mean TT ADE source}
\langle S_d \rangle = \nabla \cdot \textbf{D} \nabla \langle
A\rangle - \nabla \cdot \textbf{D} \nabla \langle E\rangle
\end{equation}
\end{subequations}

where the divergence of the advection term has been developed in
order to annihilate the fluid source and sink terms. The boundary
value problems~(\ref{eq:A BVP}) and~(\ref{eq:LE BVP}) involve that
the finite moments of the age and life expectancy pdfs exist for
homogeneous boundary conditions. By definition, the mean groundwater
age in a steady-flow reservoir, or mean residence time, can be
determined from the average of the normalized flux concentration
response to a narrow flux input uniformly applied on the recharge
limits, since this breakthrough curve corresponds to the water
molecules residence time
distribution~\cite{Danckwerts53,Kreft78,Jury90}. The mean
groundwater age can then be calculated by prescribing at all inflow
boundaries a solute mass that is proportional to the water
flux~\cite{Harvey95}. Consequently, Eqs.~(\ref{eq:mean A ADE}) and
(\ref{eq:mean LE ADE}) can be solved by assigning $\langle A\rangle
= 0$ on the inlet limits $\Gamma_-$, and $\langle E\rangle = 0$ on
the outlet limits $\Gamma_+$, respectively. Mean age and mean life
expectancy are continuously generated during groundwater flow, since
porosity acts as a source term in Eqs.~(\ref{eq:mean A ADE}) and
(\ref{eq:mean LE ADE}). This source term indicates that groundwater
is aging one unit per unit time, in average. The mean age and mean
life expectancy equations can be easily handled by numerical codes
that solve ADEs, by distributing a source term equal to porosity,
and by reversing the velocity field for the case of
Eq.~(\ref{eq:mean LE ADE}). Eq.~(\ref{eq:mean TT ADE}) would require
the boundary conditions $\langle T\rangle = \langle E\rangle$ on
$\Gamma_-$, and $\langle T\rangle = \langle A\rangle$ on $\Gamma_+$.
However, mean transit time can rather be obtained by solving the
Eqs.~(\ref{eq:mean A ADE}) and (\ref{eq:mean LE ADE}), and by
post-processing Eq.~(\ref{eq:meanTT}). If dispersion is nil
Eq.~(\ref{eq:mean TT ADE}) is simply $\textbf{q} \cdot \nabla
\langle T\rangle = 0$, the solution of which is comparable to the
stream lines in a flow model, and associates to each line the
advective transit time from inlet to outlet.

According to Parker and van Genuchten~\cite{Parker84,Parker86},
Kreft and Zuber~\cite{Kreft78,Kreft86},~and Sposito and
Barry~\cite{Sposito87}, temporal moments have a real physical
meaning if they are related to flux concentration functions, while
spatial moments must characterize resident concentration functions.
Flux and resident concentration depend themselves on the measurement
technique~\cite{Sposito87,Rubin03}. An age date as deduced from
isotopic measurements corresponds to an age average over the number
of molecules in the water sample, and may often be close to a
resident concentration. Mean age computations using
Eq.~(\ref{eq:mean A ADE}) are, therefore, well-suited for fitting
isotopic age dates.
\subsection{Theoretical illustration example.}
The numerical solutions presented in this work have been performed
using the Laplace Transform Galerkin (LTG) finite element
technique~\cite{Sudicky89}, which allows eliminating the
time-derivative in Eqs.~(\ref{eq:A ADE}) and~(\ref{eq:LE ADE}). The
Cauchy type boundary conditions ~(\ref{eq:BC1 A ADE})
and~(\ref{eq:BC1 LE ADE}) imply that Eqs.~(\ref{eq:A ADE})
and~(\ref{eq:LE ADE}) must be formulated according to their
divergence form~\cite{Diersch98}, such that a total mass flux
continuity at the boundaries must be properly handled. The assembled
linear system of equations is solved by the accurate ILUT-based
sparse iterative solver~\cite{Saad94} with complex arguments and
GMRES~\cite{Saad86} or BiCGSTAB~\cite{Vorst92} acceleration. The
numerical inversion of the Laplace transformed functions is
performed by the algorithm of Crump~\cite{Crump76}. The
quotient-difference algorithm is used to accelerate the convergence
of the Fourier series~\cite{Hoog82}. This algorithm proved to be
very efficient to treat inversion at the neighborhood of
discontinuities or sharp fronts, and the required computational
effort, which is linearly proportional to the $2n + 1$ number of
discrete Laplace variables, is highly diminished with respect to
other acceleration methods.

\begin{figure}[tbp]
\begin{center}
\includegraphics[width=\textwidth]{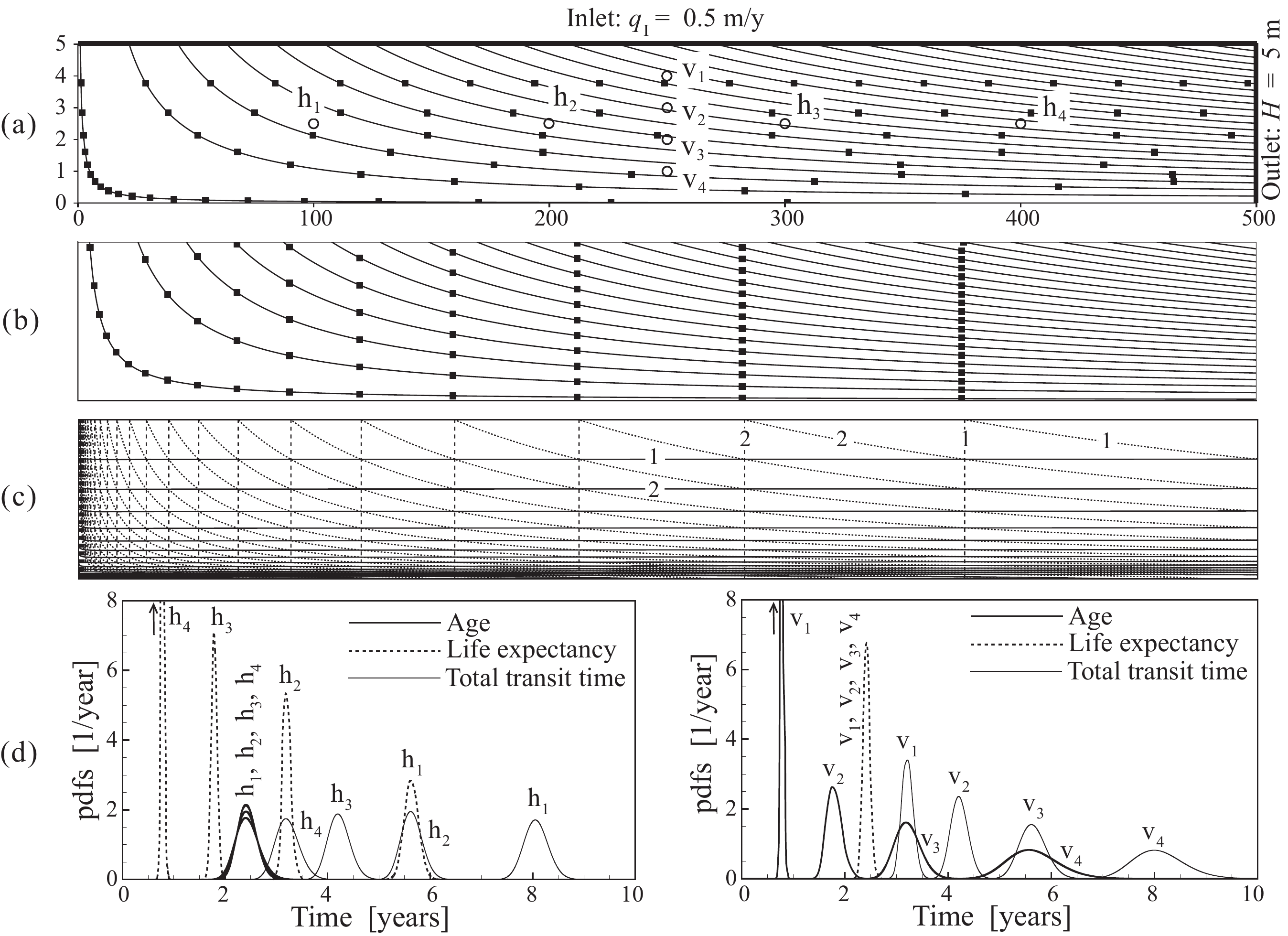}
\end{center}
\caption{Age, life expectancy and transit time computations in a 2-D
vertical theoretical aquifer. (a) Geometry, boundary conditions and
flow field representation by forward particle-tracking with
marker-isochrones each 1~year; (b) Backward particle-tracking with
marker-isochrones each 1~year; (c) Spatial distribution of mean age
(solid lines), mean life expectancy (dashed lines) and mean transit
time (dotted lines) with 1~year of time-increment; (d) Age, life
expectancy and transit time pdfs at some observations points.
Parameters: $K = 10^{-4}$~m/s, $\phi = 0.35$, $\alpha_L = 0.1$~m,
$\alpha_T = 0.001$~m, $D_\textrm{m} = 0$.\label{fig:F3}}
\end{figure}

We use here the simple example of a theoretical vertical aquifer
section (see Fig.~\ref{fig:F3}) to illustrate age, life expectancy
and transit time computations. The configuration of the flow problem
corresponds to the typical case known as 'well-mixed reservoirs',
that generate an exponential-like transit time distribution at
outlet. This aspect will be discussed in more details in
Section~\ref{sec:charactimes}. The aquifer is homogeneous, and is
uniformly recharged on top by a constant infiltration rate. The
outlet, which could represent a trench, is simulated by means of a
prescribed constant hydraulic head from the top to the bottom of the
section. We solve the boundary value problems~(\ref{eq:A BVP})
and~(\ref{eq:LE BVP}), as well as Eqs.~(\ref{eq:mean A ADE})
and~(\ref{eq:mean LE ADE}). The model is discretized into 100'000
bilinear quadrangles of size $\Delta x = 0.5$~m and $\Delta z =
0.25$~m. Fig.~\ref{fig:F3}a shows the flow field by means of a
forward particle-tracking representation. Fig.~\ref{fig:F3}b gives
the backward particle-tracking solution, which represents the purely
advective life expectancy spatial distribution. In
Fig.~\ref{fig:F3}c, we have plotted the solutions of
Eqs.~(\ref{eq:mean A ADE}),~(\ref{eq:mean LE ADE})
and~(\ref{eq:meanTT}). Mean age and mean life expectancy are in very
good agreement with particle-tracking solutions. Isolines of mean
age are horizontally distributed in relation to the flow
configuration: velocity increases upstream to downstream (increase
of hydraulic gradient towards outlet), but also becomes more and
more horizontally distributed. Travel paths to reach a same depth
are longer downstream than upstream, but since velocity increases
downstream, it creates a compensating effect and the corresponding
travel times finally become similar, in average. Mean life
expectancy gradients are mainly horizontal, and gradually increasing
upstream. Mean transit time provides a good representation of the
flow field; its distribution is very close to the flow lines shown
in Fig.~\ref{fig:F3}a. However, the mean transit time solution owns
the additional information on the total time required by water
molecules to travel from inlet to outlet. Finally, in
Fig.~\ref{fig:F3}d are given some observed pdfs (see
Fig.~\ref{fig:F3}a for the location of the observation points).
Successive age pdfs along a horizontal profile confirm the vertical
gradient of mean age shown in Fig.~\ref{fig:F3}c, each age pdf at a
same depth being very comparable to one other. Along a vertical
profile, successive age pdfs are more and more spread in relation to
the effect of longitudinal dispersion, which is bigger the longer
the travel path. The behavior of the life expectancy pdfs is similar
to that of age, but in the reversed direction of velocity. The
transit time pdfs along a horizontal profile are very similar to
each other; they are simply shifted along the axis of time. This
reveals a same amount of dispersion affecting transport of water
molecules along each travel path from inlet to outlet. Vertically,
the functions $g_T(\textit{\textbf{x}},t)$ show more and more
dispersion with increasing depth, in relation to the vertical mixing
of water molecules.
\section{Generalized Reservoir Theory.}
Recent studies~\cite{Etcheverry99,Etcheverry00} proposed a direct
approach to calculate outlet transit time distributions by applying
the reservoir theory (RT)~\cite{Eriksson71,Bolin73} to the average
groundwater age field, resulting from a pure advective transport
solution. In the following, we show how the RT can be generalized to
advective-dispersive systems by manipulating the forward and
backward transport equations. We first introduce some characteristic
reservoir time probability distributions. Two equivalent
mathematical formulations are then proposed, in order to evaluate
the discharge zone transit time pdf as a function of the reservoir
internal physical properties.
\subsection{Characteristic reservoir distributions.}
When a specific age distribution is assigned to each elementary
water volume in the reservoir, the volume of mobile water can be
classified in a cumulative manner with respect to the age occurrence
in the reservoir. Let $M_A(t)$ be the cumulated amount of water
molecules in the reservoir with an age inferior or equal to a
particular value \textit{t}, and $m_A(t)$ be the corresponding
probability function, or reservoir internal age cdf. The function
$m_A(t)$ is the porous volume normalized function $M_A(t)$, which is
evaluated by integrating over $\Omega$ the probability of finding
water molecules with an age \textit{t} or less, assuming that each
molecule has entered the system on $\Gamma_-$:
\begin{equation}\label{eq:intAcdf}
m_A (t) = \frac{{M_A (t)}}{{M_0 }} = \frac{1}{{M_0 }}\int_\Omega
{\phi G_A (\textit{\textbf{x}},t)}d\Omega = \frac{1}{{M_0
}}\int_\Omega  {\phi \left( {\int_0^t {g_A (\textit{\textbf{x}},\tau
) d\tau } } \right)}d\Omega
\end{equation}

with $M_0$ being the total amount of mobile water (aquifer porous
volume). The function $m_A(t)$ cumulates the probability of finding
water molecules which have travelled until a position
\textit{\textbf{x}} before time \textit{t}. The function $M_A(t)$ is
zero at origin and tends towards the total porous volume $M_0$ at
infinity. The internal age frequency distribution function
$\psi_A(t)$ is the pdf associated to $m_A(t)$, and from
Eq.~(\ref{eq:PDFdef}) it follows
\begin{equation}\label{eq:intApdf}
\psi_A (t) = \frac{{\partial m_A (t)}}{{\partial t}} = \frac{1}{{M_0
}} \int_\Omega  {\phi g_A (\textit{\textbf{x}},t)d\Omega }
\end{equation}

Note that Eq.~(\ref{eq:intApdf}) corresponds to the zeroth spatial
moment of the age pdf, which is equivalent to the age
mass~\cite{Rubin03}. The function $\psi_A(t)$ gives the probability
density of finding elements in $\Omega$ that have reached the age
\textit{t}, and $\psi_A(t)dt$ is the probability that \textit{A}
lies in the interval $[t, t + dt]$. Thus, the quantity $M_0
\psi_A(t)$ represents the number of elements per unit time (or flow
rate fraction) that are in the interval $[t, t + dt]$, and is
equivalent to the zeroth spatial moment of the age pdf
$g_A(\textit{\textbf{x}},t)$. We similarly define the internal life
expectancy cdf $m_E(t)$ and pdf $\psi_E(t)$:
\begin{equation}\label{eq:intLEcdf}
m_E (t) = \frac{{M_E (t)}}{{M_0 }} = \frac{1}{{M_0 }}\int_\Omega
{\phi G_E (\textit{\textbf{x}},t)} d\Omega = \frac{1}{{M_0
}}\int_\Omega  {\phi \left( {\int_0^t {g_E (\textit{\textbf{x}},\tau
) d\tau } } \right)} d\Omega
\end{equation}
\begin{equation}\label{eq:intLEpdf}
\psi_E (t) = \frac{{\partial m_E (t)}}{{\partial t}} = \frac{1}{{M_0
}}\int_\Omega  {\phi g_E (\textit{\textbf{x}},t)d\Omega }
\end{equation}

The function $m_E(t)$ cumulates the probability that a water
molecule in the reservoir will reach the outlet before time
\textit{t}.

We finally consider the internal distribution of the groundwater
molecules transit time pdf $g_T$ as deduced from the convolution
integral in Eq.~(\ref{eq:TT PDF}), to introduce the functions
$m_T(t)$ and $\Psi(t)$ as the internal transit time cdf and pdf,
respectively:
\begin{equation}\label{eq:intTTcdf}
m_T (t) = \frac{{M_T (t)}}{{M_0 }} = \frac{1}{{M_0 }}\int_\Omega
{\phi G_T (\textit{\textbf{x}},t)d\Omega }  = \frac{1}{{M_0
}}\int_\Omega {\phi \left( {\int_0^t {g_T (\textit{\textbf{x}},\tau
)d\tau } } \right)d\Omega }
\end{equation}
\begin{equation}\label{eq:intTTpdf}
\Psi (t) = \frac{{\partial m_T (t)}}{{\partial t}} =
\frac{1}{{M_0}}\int_\Omega  {\phi g_T (\textit{\textbf{x}},t)d\Omega
}
\end{equation}

The function $M_T(t)$ corresponds to the mass of mobile water in the
reservoir having a transit time inferior or equal to \textit{t}. The
function $\Psi(t)$ characterizes the probability density of finding
water molecules within the reservoir that have a transit time
inferior or equal to $t$, and the quantity $M_0 \Psi(t)dt$  gives
the amount of water molecules that travel through $\Omega$ within
the time interval $[t, t + dt]$.
\subsection{The transit time pdf of discharge and recharge zones.}
A reservoir discharge zone is a particular portion of finite size,
which intercepts the groundwater molecules that contribute to the
outflow rate. These water molecules have contrasted arrival times
that converging water fluxes mix together before flowing out. At a
given position on a discharge boundary, the transit time pdf is the
age pdf. To characterize the contribution of each age flux event on
an outflow boundary in terms of transit time probability, it is
convenient to average the age probability fluxes on $\Gamma_+$.
Under steady-flow regime, the representative transit time
distribution $\varphi_A(t)$ of the reservoir outlet zone can be
defined as a flux averaged concentration~\cite{Rubin03}, i.e.
$\varphi_A(t)$ is evaluated as the flow rate-normalized sum on
$\Gamma_+$ of the total age mass flux response function
$\textbf{J}_A$ resulting from a unit flux impulse on $\Gamma_-$:
\begin{equation}\label{eq:phiAdef1}
\varphi _A (t) = \frac{1}{{F_0 }} \int_{\Gamma _ +  } {{\bf{J}}_A
\cdot \textit{\textbf{n}}\,d\Gamma }  = \frac{1}{{F_0 }}
\int_{\Gamma _ +  } {[{\bf{q}}g_A  - {\bf{D}}\nabla g_A ] \cdot
\textit{\textbf{n}}\,d\Gamma }
\end{equation}

where \textit{\textbf{n}} is a normal unit vector pointing outward
the boundary, and where $F_0$ represents the total discharge flow
rate throughout the bounded domain, which at steady-state is
evaluated by:
\begin{equation}\label{eq:F0}
F_0 = \int_{\Gamma _ +  } {{\bf{q}}(\textit{\textbf{x}}) \cdot
\textit{\textbf{n}}\,d\Gamma } = \int_{\Gamma _ -  }
{{\bf{q}}(\textit{\textbf{x}}) \cdot \textit{\textbf{n}}\,d\Gamma }
\end{equation}

Note that for the sake of simplicity, internal sources and sinks are
neglected, $q_\textrm{I} = q_\textrm{O} = 0$. While flux
concentrations are defined with respect to a control plane
orthogonal to the velocity vector direction, the outlet transit time
probability function in Eq.~(\ref{eq:phiAdef1}) is defined by the
projection of the total age flux on the arbitrary-shaped boundary
$\Gamma_+$. Inserting Eq.~(\ref{eq:A flux conc}) into
Eq.~(\ref{eq:phiAdef1}) produces the equivalent relation:
\begin{equation}\label{eq:phiAdef2}
\varphi _A (t) = \frac{1}{{F_0 }} \int_{\Gamma _ +  } {\left[
{{\bf{q}}g_A^f - \left( {{\bf{I}} - \frac{{{\bf{q}} \otimes
{\bf{q}}}}{{\left\| {\bf{q}} \right\|^2 }}} \right){\bf{D}}\nabla
g_A } \right] \cdot \textit{\textbf{n}}\,d\Gamma }
\end{equation}

With Eq.~(\ref{eq:phiAdef2}), one can see that, if the velocity
vector \textbf{q} points in the direction of the outward normal unit
vector \textit{\textbf{n}}, then the dispersive term inside the
brackets vanishes, and the pdf $\varphi_A(t)$ is equal to the total
steady flow rate-normalized sum of the flux-weighted age flux
concentration pdfs on the outlet limit. This is always the case in
one-dimension. The cross product term in Eq.~(\ref{eq:phiAdef2})
reveals also that since the outflow limit is of arbitrary shape,
then when velocity does not point in the direction of
\textit{\textbf{n}} a dispersive correction term is required. This
is related to the fact that flux concentrations are defined with
respect to a control plane which is orthogonal to the velocity
direction.

The discharge flow rate can also be described as a cumulative
function of the transit time values. We introduce the function
$f_A(t)$ as the transit time cdf of the reservoir outlet, i.e.
$f_A(t)$ is the probability that the molecules flow out with a
transit time \textit{t} or less, such that it corresponds to the
normalized cumulated outflow function $F_A(t)$:
\begin{equation}\label{eq:fA}
f_A (t) = \frac{{F_A (t)}}{{F_0}} = \int_0^t {\varphi _A (\tau )
d\tau }
\end{equation}

Note that the function $f_A(t)$ can also be viewed as the integral
on $\Gamma_+$ of the total mass flux deduced from the resident
concentration solutions of the ADE~(\ref{eq:A ADE}) for a unit
step-input of mass flux on $\Gamma_-$. By analogy, the life
expectancy pdf and cdf of the reservoir inlet $\Gamma_-$ can
similarly be defined by
\begin{equation}\label{eq:phiEdef}
\begin{split}
\varphi _E (t) & = \frac{1}{{F_0}} \int_{\Gamma _ -  } {{\bf{J}}_E
\cdot \textit{\textbf{n}}\,d\Gamma } = - \frac{1}{{F_0
}}\int_{\Gamma _ - }{[{\bf{q}}g_E + {\bf{D}}\nabla g_E ]
\cdot \textit{\textbf{n}}\,d\Gamma } \\
               & = - \frac{1}{{F_0}} \int_{\Gamma _ -  }
{\left[ {{\bf{q}}g_E^f + \left( {{\bf{I}} - \frac{{{\bf{q}} \otimes
{\bf{q}}}}{{\left\| {\bf{q}} \right\|^2 }}} \right){\bf{D}}\nabla
g_E } \right] \cdot \textit{\textbf{n}}\,d\Gamma }
\end{split}
\end{equation}
\begin{equation}\label{eq:fE}
f_E (t) = \frac{{F_E (t)}}{{F_0}} = \int_0^t {\varphi _E (\tau )
d\tau }
\end{equation}

Eq.~(\ref{eq:TT flux conc}) indicates that the transit time flux pdf
equals both the life expectancy flux pdf at inlet, and the age flux
pdf at outlet. As a matter of fact, assuming neither addition nor
subtraction of mass during transport, steady-flow conditions and
stationary state, then, theoretically, every probability flux
$\textbf{J}_A$ on $\Gamma_+$ has an identical counterpart
$\textbf{J}_E$ on $\Gamma_-$ and vice-versa. Therefore,
Eqs.~(\ref{eq:phiAdef1}),~(\ref{eq:phiAdef2}) and~(\ref{eq:phiEdef})
relate to the same and unique function, $\varphi(t) = \varphi_A(t) =
\varphi_E(t)$, and thus $f(t) = f_A(t) = f_E(t)$. In fact, each
element added in the reservoir at the inlet must exit at some
position $\textit{\textbf{x}}_o$ of the outlet sooner or later. Each
element added at the position $\textit{\textbf{x}}_o$ at the outlet
must travel the same distance upstream, and thus spend the same
time-span within the reservoir, backward in time, before reaching a
position somewhere at the inlet limit.

At the reservoir outlet, the arrival times $g_A$ are distributed
along the discharge boundary, implying mixing and superposition of
the information carried by each breakthrough curve. Moreover, within
the reservoir mixing processes can also be important, and the true
minimum and maximum ages can be diluted. To characterize an outlet
representative transit time distribution, we must ensure that the
minimum and the maximum ages are captured. Technical problems can
often occur when solving Eq.~(\ref{eq:phiAdef1}) or
Eq.~(\ref{eq:phiEdef}), because mass flux line/surface integration
is required. If the outlet is of small size, then the capture of
calculated breakthrough curves, or the identification of particle
arrivals, reveals to be ill-posed, mainly because of the loss of
information due to the mixing of converging fluxes. Hence, numerical
methods will generally require a high level of refinement in the
neighborhood of these integration limits, which rapidly becomes a
computational limiting factor. Because the transit time pdf
$\varphi(t)$ on the inlet limit is identical on the outlet limit,
discretization methods imply that the number of observation nodes
should be the same at the inlet and at the outlet, in order to be
able to recover the same breakthrough curves. In other words, the
temporal resolution of the curve, when a counting of the individual
arrival times is performed, is a direct function of the spatial
refinement in the vicinity of exit zones. The same restriction
affects other simulation methods, such as the random-walk procedure.

In the following, we propose an alternative approach that is relaxed
from the above-mentioned practical problems. Eqs.~(\ref{eq:A BVP})
and~(\ref{eq:LE BVP}) are considered to simulate the age and life
expectancy probability distributions in the reservoir $\Omega$.
Integrating Eq.~(\ref{eq:A ADE}) over $\Omega$, and making use of
the divergence theorem ($ \int_\Omega {\nabla \cdot {\bf{F}} d\Omega
}  = \int_\Gamma {{\bf{F}} \cdot\textit{\textbf{n}}\,d\Gamma } $)
results in
\begin{equation}\label{eq:AADEint1}
\int_{\Gamma _ +} {[{\bf{q}}g_A  - {\bf{D}}\nabla g_A] \cdot
\textit{\textbf{n}}\,d\Gamma } + \frac{\partial }{{\partial
t}}\int_\Omega {\phi g_A d\Omega} = - \int_{\Gamma _ -}
{[{\bf{q}}g_A  - {\bf{D}}\nabla g_A ] \cdot \textit{\textbf{n}}\,
d\Gamma }
\end{equation}

Normalizing Eq.~(\ref{eq:AADEint1}) by the steady flow rate $F_0$
and accounting for Eq.~(\ref{eq:intApdf}) and~(\ref{eq:phiAdef1}),
Eq.~(\ref{eq:AADEint1}) can be turned into the following form:
\begin{equation}\label{eq:AADEint2}
\varphi _A (t) + \tau _0 \frac{{\partial \psi _A (t)}}{{\partial
 t}} = - \frac{1}{{F_0 }}\int_{\Gamma _ -}
{[{\bf{q}}g_A  - {\bf{D}}\nabla g_A ] \cdot \textit{\textbf{n}}\,
d\Gamma }
\end{equation}

with the quantity $\tau_0$ being the turnover time commonly defined
at steady-state as the ratio of porous volume to flow rate:
\begin{equation}\label{eq:tau0}
\tau _0  = \frac{{M_0 }}{{F_0 }}
\end{equation}

Similarly, the integration of the ADE~(\ref{eq:LE ADE}) has the form
\begin{equation}\label{eq:LEADEint1}
- \int_{\Gamma _ +  } {[{\bf{q}}g_E + {\bf{D}}\nabla g_E ] \cdot
\textit{\textbf{n}}\,d\Gamma } + \frac{\partial }{{\partial
t}}\int_\Omega {\phi g_E d\Omega } = \int_{\Gamma _ -} {[{\bf{q}}g_E
- {\bf{D}}\nabla g_E] \cdot \textit{\textbf{n}}\, d\Gamma }
\end{equation}

Normalizing Eq.~(\ref{eq:LEADEint1}) by $F_0$ and accounting for
Eq.~(\ref{eq:intLEpdf}) and~(\ref{eq:phiEdef}) yields
\begin{equation}\label{eq:LEADEint2}
\varphi _E (t) + \tau _0 \frac{{\partial \psi _E (t)}}{{\partial
 t}} = \frac{1}{{F_0 }}\int_{\Gamma _ -}
{[{\bf{q}}g_E - {\bf{D}}\nabla g_E ] \cdot \textit{\textbf{n}}\,
d\Gamma }
\end{equation}

The boundary integrals in the right-hand sides of
Eqs.~(\ref{eq:AADEint2}) and~(\ref{eq:LEADEint2}) can be simplified
by accounting for the boundary conditions~(\ref{eq:BC1 A ADE})
and~(\ref{eq:BC1 LE ADE}). For instance, inserting the boundary
condition~(\ref{eq:BC1 A ADE}) into Eq.~(\ref{eq:AADEint2}) reduces
the boundary integral to $- F_0 \delta(t)$, and inserting the
boundary condition~(\ref{eq:BC1 LE ADE}) into
Eq.~(\ref{eq:LEADEint2}) reduces the boundary integral to $F_0
\delta(t)$, and Eqs.~(\ref{eq:AADEint2}) and~(\ref{eq:LEADEint2})
become:
\begin{subequations}
\begin{equation}\label{eq:GRTAtmp}
\varphi _A (t) + \tau _0 \frac{{\partial \psi _A (t)}}{{\partial t}}
= \delta (t)
\end{equation}
\begin{equation}\label{eq:GRTEtmp}
\varphi _E (t) + \tau _0 \frac{{\partial \psi _E (t)}}{{\partial t}}
= \delta (t)
\end{equation}
\end{subequations}

With Eq.~(\ref{eq:GRTAtmp}) we have recovered the RT formulation of
Eriksson~\cite{Eriksson71}, in which the outlet zone transit time
pdf $\varphi_A(t)$ is proportional to the first derivative of the
internal age pdf $\psi_A(t)$, thus characterizing the probability
for the water molecules of being removed from $\Omega$ per unit
time. Eq.~(\ref{eq:GRTEtmp}) is an equivalent formulation which
relates the recharge boundary life expectancy distribution to the
internal life expectancy distribution. Since the functions
$\varphi_A(t)$ and $\varphi_E(t)$ are equal, it follows from
Eqs.~(\ref{eq:GRTAtmp}) and~(\ref{eq:GRTEtmp}) that $\psi_A(t) =
\psi_E(t)$, which allows writing the following general RT
formulation:
\begin{equation}\label{eq:GRT}
\varphi (t) + \tau _0 \frac{{\partial \psi (t)}}{{\partial t}} =
\delta (t)
\end{equation}

Eq.~(\ref{eq:GRT}) generalizes the RT to advective-dispersive solute
transport processes, and is valid for both age and life expectancy.
If dispersion is set to zero, the function $\psi(t)$ can be
evaluated by integration of the field $g_A(\textit{\textbf{x}},t) =
\delta(t - \langle A\rangle(\textit{\textbf{x}}))$ as proposed by
Etcheverry and Perrochet~\cite{Etcheverry99,Etcheverry00}.
Similarly, the function $\psi(t)$ can be evaluated by integration of
the field $g_E(\textit{\textbf{x}},t) = \delta(t - \langle
E\rangle(\textit{\textbf{x}}))$. Since $\psi(t) = \psi_A(t) =
\psi_E(t)$, it follows from Eq.~(\ref{eq:CDFdef}) that $M(t) =
M_A(t) = M_E(t)$. This points out the importance in allowing age and
life expectancy dispersive fluxes crossing naturally the outlet and
inlet boundary portions, because the use of the homogeneous Neumann
condition on $\Gamma_+$ and may lead to different results for
$\varphi(t)$ if contrasted boundary configurations and flow
conditions exist. The third-kind boundary conditions~(\ref{eq:BC1 A
ADE}) and~(\ref{eq:BC1 LE ADE}) are the most meaningful conditions
for solving the age and life expectancy problems. They ensure a pure
total flux pulse input entering the system at inlet for the age
problem, and at outlet for the life expectancy problem. These
conditions become homogeneous for $t > 0$ (zero flux), and do not
allow backward mass losses by dispersion. This would not be the case
when using a Dirichlet type condition, which may lead to incorrect
solute mass balances. For the one-dimensional case, the use of the
Dirichlet condition permits the simulation of the age and life
expectancy pdfs, but directly for flux concentration pdfs (see
Appendix~A).

The fundamental relation between the outlet transit time cdf $f(t)$
and the internal age pdf $\psi(t)$ given by the RT is obtained after
integration of Eq.~(\ref{eq:GRT}):
\begin{equation}\label{eq:GRTmassbal1}
F_0 - F(t) = M_0 \psi (t) = \frac{{\partial M(t)}}{{\partial t}}
\end{equation}
or
\begin{equation}\label{eq:GRTmassbal2}
f(t) + \tau _0 \psi (t) = 1
\end{equation}

Eq.~(\ref{eq:GRTmassbal1}) indicates that the outflow of water
molecules that leave the system through $\Gamma_+$ with an age older
than \textit{t} is balanced by the number of elements per unit time
within $\Omega$ that reach the age \textit{t}, i.e. in the interval
$[t,t + dt]$. The corresponding volume of groundwater reaching the
age \textit{t} in $\Omega$ is $M_0 \psi(t) dt$. In other words, the
flow rate fraction of age \textit{t} or less at the outlet is a
function of the spatial occurrence in the reservoir of water
molecules of age \textit{t} or less. Since the function $f(t)$ is
zero at the origin, it follows that the value of $\psi(t)$ at origin
is the turnover rate $\sigma_0$ ($\psi(0) = \sigma_0 = 1/\tau_0$),
independently of the level of dispersion. Since $f(t)$ is a
cumulative function, then $\psi(t)$ must be monotonically decreasing
and $M(t)$ must be an increasing function with monotonically
decreasing increments~\cite{Eriksson71}. The pdf $\psi(t)$ is
constant from zero to the minimum age $t_\textrm{min}$ at outlet,
with $\psi(0,\ldots,t_\textrm{min}) = \sigma_0$, which throws light
on the fact that the probability per unit time of finding elements
in $\Omega$ that have reached the age $\textit{t} \leq
t_\textrm{min}$ is certain. From Eq.~(\ref{eq:GRTmassbal1}) it
follows that $M(t)$ has a constant derivative equal to $M_0 \psi(0)
= F_0$ until the minimum age $t_\textrm{min}$ is reached at outlet.
Note that the same considerations can be made for $f(t) = f_E(t)$,
$\psi(t) = \psi_E(t)$ and $M(t) = M_E(t)$.

\begin{figure}[tbp]
\begin{center}
\includegraphics[width=\textwidth]{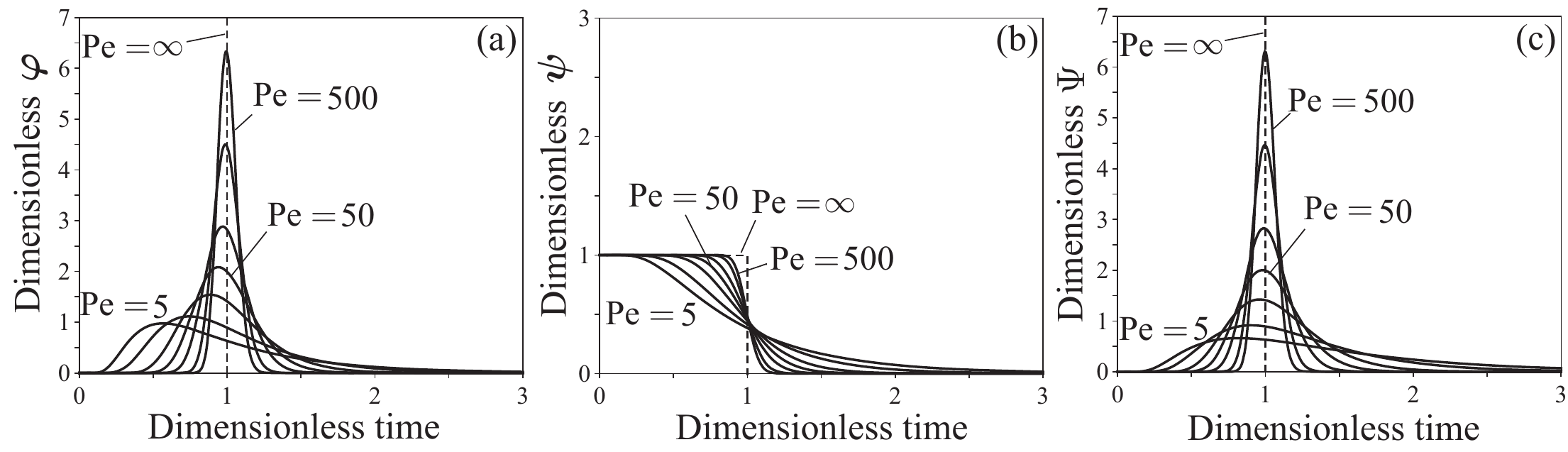}
\end{center}
\caption{Reservoir theory pdfs for a 1-D semi-infinite flow domain,
as a function of the P\'{e}clet number for $\textrm{Pe} = 5, 10, 25,
50, 100, 250~\textrm{and}~500$. (a) Outlet transit time pdf; (b)
Internal age (or life expectancy) pdf; (c) Internal transit time
pdf. Time is normalized by the average turnover time $\tau_0 = L/v$,
\textit{x} by a characteristic length \textit{L}, and $\textrm{Pe} =
Lv/D$.\label{fig:F4}}
\end{figure}

Fig.~\ref{fig:F4} illustrates in 1-D the outlet (or inlet) transit
time pdf, the internal age (or life expectancy) pdf, and the
internal transit time pdf resulting from the analytical resolution
of Eqs.~(\ref{eq:intApdf}),~(\ref{eq:intTTpdf}) and~(\ref{eq:GRT})
(see Appendix~A). Under pure convective transport conditions (dashed
lines in Fig.~\ref{fig:F4}), the function $\varphi(t)$ equals the
piston-flow transit time pdf $\delta(t - \tau_0)$, and the function
$\psi(t)$ is the Heaviside function $\textrm{H}(\tau_0 - t)/\tau_0$,
such that $\psi(t) = \psi(0) = 1/\tau_0$ until $t_\textrm{min} =
\tau_0$, and $\psi(t) = 0$ after $t_\textrm{min}$. The first
temporal moment of the pdf $\varphi(t)$ (average transit time at
outlet $\tau_\textrm{t}$) is dispersion-independent, and equals the
turnover time $\tau_0$. The average internal transit time
$\tau_\textrm{it}$ and the average internal age $\tau_\textrm{i}$
(first temporal moments of $\Psi(t)$ and $\psi(t)$) are dispersion
dependent. Increasing longitudinal dispersion (low P\'{e}clet
numbers) generates short arrival times and tailing effects (see the
variances of the pdfs in Appendix~A), and thus old arrival times at
the outlet as well as old ages within the domain, which are visible
on the three functions $\varphi(t)$, $\psi(t)$ and $\Psi(t)$ for a
range of P\'{e}clet numbers. The function $\psi(t)$ is constant from
0 until the minimum age at outlet. Since the Cauchy type condition
prevents backward losses by dispersion at $\textit{x} = 0$, the
value of $\psi(t)$ at the origin is always $1/\tau_0$ for any
P\'{e}clet number (Fig.~\ref{fig:F4}b).
From the spatial organization of age or life expectancy occurrence
it is possible to predict the transit time distribution of a
reservoir outlet (or equivalently the life expectancy distribution
of a reservoir inlet), without the need of 'counting' the arrivals
of the water molecules at a boundary of finite size. Thus, the pdf
$\varphi(t)$ defined in Eqs.~(\ref{eq:phiAdef1})
or~(\ref{eq:phiEdef}) as a pure boundary property becomes a property
of the reservoir internal structure and hydro-dispersive
characteristics. The information that can be lost when $\varphi(t)$
is directly evaluated at the reservoir exit zone (or inlet zone) is
recovered with Eq.~(\ref{eq:GRT}). A far more accurate evaluation of
$\varphi(t)$ is thus achieved, for which the main operation (domain
integrals~(\ref{eq:intApdf}) and~(\ref{eq:intLEpdf})) is not
time-consuming and can easily be implemented for one-, two- and
three-dimensional systems. A convenient way to compute
Eq.~(\ref{eq:GRT}) is to work in the Laplace space, since it allows
handling all time-dependent quantities in a quasi-analytical way.
\subsection{Temporal moments of the reservoir theory probability
            density functions.}
A direct consequence of the RT is that the expected value of the
mean residence time, i.e. the average transit time $\tau_\textrm{t}$
at outlet or inlet, equals the reservoir mean turnover time
$\tau_0$. This property can be found by calculating the first
temporal moment of the transit time pdf $\varphi(t)$, and by making
use of Eq.~(\ref{eq:GRTmassbal2}):
\begin{equation}\label{eq:m1phi}
\tau _{\rm{t}} = \int_0^{ + \infty } {t \varphi (t) dt} = \int_0^{ +
\infty } {[ 1 - f(t) ] dt} = \tau _0 \int_0^{ + \infty } { \psi (t)
dt} = \tau _0 = \frac{M_0}{F_0}
\end{equation}
where use has been made of the pdf property in
Eq.~(\ref{eq:PDFdef}). Since the reservoir is considered under
steady flow conditions, internal average time characteristics can be
defined. The temporal moments of the global functions $\psi(t)$ and
$\Psi(t)$ have physical significance since resident concentration
has been integrated in space. The mean internal age
$\tau_\textrm{ia}$ and the mean internal life expectancy
$\tau_\textrm{ie}$ are deduced from the first temporal moment of the
function $\psi(t)$. Integrating by parts and making use of the
relation~(\ref{eq:GRT}) results in:
\begin{eqnarray}\label{eq:m1psi}
\tau _{\rm{i}} & = & \int_0^{ + \infty } {t \psi (t)dt} = \left.
{\frac{{t^2 \psi (t)}}{2}} \right|_0^{+\infty} - \frac{1}{2}\int_0^{
+\infty } {t^2 \frac{\partial \psi (t)}{\partial t} dt} = \frac{1}{2
\tau_0} \int_0^{
+\infty } {t^2 \varphi(t) dt}\nonumber\\
& = & \frac{\mu_{\rm{2}} [\varphi]}{2 \tau_0} = \frac{\tau_0}{2} (1
+ \frac{\sigma^2 [\varphi]}{\tau^2_0})
\end{eqnarray}

Since $\psi_A(t) = \psi_E(t) = \psi(t)$, we have set
$\tau_\textrm{ia} = \tau_\textrm{ie} = \tau_\textrm{i}$. Using
Eq.~(\ref{eq:meanTT}), the average internal transit time
$\tau_\textrm{it}$ is deduced from the first temporal moment of the
function $\Psi(t)$, and is found to be equal to twice the value of
$\tau_\textrm{i}$:
\begin{eqnarray}\label{eq:m1psiT}
\tau _{{\rm{it}}} & = & \int_0^{ + \infty } {t \Psi (t)dt} =
\frac{1}{{M_0 }} \int_\Omega  {\phi \int_0^{ + \infty } { t
g_T(\textit{\textbf{x}},\textit{t})dt} d\Omega } = \frac{1}{{M_0 }}
\int_\Omega  {\phi \langle T\rangle d\Omega } \nonumber\\
& = & \frac{1}{{M_0 }} \int_\Omega  {\phi \langle A\rangle d\Omega }
+ \frac{1}{{M_0 }} \int_\Omega  {\phi \langle E\rangle d\Omega } =
2\tau _{\rm{i}}
\end{eqnarray}

Using Eq.~(\ref{eq:GRT}), integrating by parts and inserting
Eq.~(\ref{eq:m1psi}), the second temporal moment and the variance of
the distribution $\varphi(t)$ reveal themselves as being functions
of $\tau_0$ and $\tau_\textrm{i}$ (or $\tau_\textrm{it}$) only:
\begin{subequations}
\begin{equation}\label{eq:m2phi}
\mu_{\rm{2}} [\varphi ] = \int_0^{ + \infty } {t^2 \varphi (t)dt} =
2\tau _0 \tau_{\rm{i}} = \tau_0 \tau_{\rm{it}}
\end{equation}
\begin{equation}\label{eq:stdphi}
\sigma^2 [\varphi] = \tau_0 (2\tau_{\rm{i}}  - \tau_0) = \tau _0
(\tau_{\rm{it}} - \tau_0)
\end{equation}
\end{subequations}

Finally note that by using the same technique than in
Eq.~(\ref{eq:m1psi}), one can show than the second moment of
$\psi(t)$ is a function of the third moment of $\varphi(t)$,
$\mu_{\rm{3}} [\varphi ] = 3 \tau_0 \mu_{\rm{2}} [\psi ]$.
\subsection{Accuracy of the RT approach.}
\begin{figure}[tbp]
\begin{center}
\includegraphics[width=\textwidth]{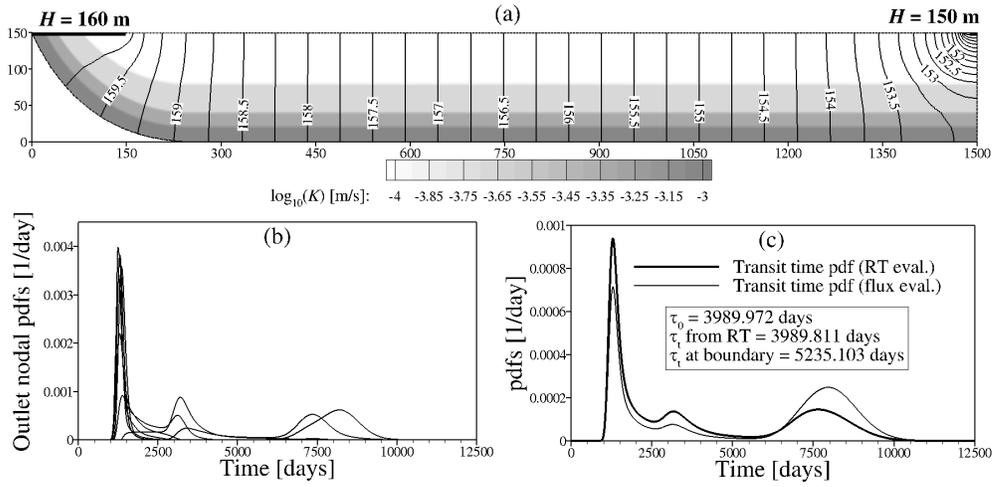}
\end{center}
\caption{Theoretical four-layered aquifer illustrating the accuracy
of the RT compared to the classical direct evaluation at the outlet
of the function $\varphi(t)$. (a) Model geometry, boundary
conditions and head solution in meters; (b) Age pdfs monitored at
outlet, from which a flux-weighted average evaluation of the transit
time pdf is performed; (c) Outlet transit time pdf evaluated at
outlet and using the RT. Transport parameters: $\alpha_L =
2.5~\textrm{m}, \alpha_T = 0.01 \alpha_L, D_\textrm{m} =
0$.\label{fig:F5}}
\end{figure}

The accuracy of the RT method compared to a classical evaluation of
the transit time pdf $\varphi(t)$ at the outlet limit is illustrated
in Fig.~\ref{fig:F5}, using a four-layered vertical aquifer. The
outlet zone is of very small size (Fig.~\ref{fig:F5}a), which forces
the information on age to be highly mixed. As attested by
Fig.~\ref{fig:F5}b, the individual age pdfs at outlet can be of very
different shape. The differences between the two evaluation methods
are very important. The flux-weighted sum of the age mass fluxes
monitored at the outlet nodes suffers from a loss of information
induced by the mixing of converging fluxes at the outlet
surroundings. With the RT, this information is recovered, since we
ensure that each contribution to the outflow rate is accounted for.
According to Eq.~(\ref{eq:m1phi}), the average residence time
$\tau_\textrm{t}$ must equal the turnover time $\tau_0$; it is clear
with this example that this property is not satisfied by straight
application of Eq.~(\ref{eq:phiAdef1}), while the RT provides a very
accurate solution (Fig.~\ref{fig:F5}c).
\section{Direct evaluation of aquifer water volumes versus age,
         life expectancy and transit time.}
The specific groundwater volumes related to a given range of ages or
residence times are important quantities to consider when addressing
aquifer management strategies. Assessing the long-term evolution of
groundwater chemistry, or defining corrective measures aiming at
restoring groundwater quality after a contamination event, indeed
requires appropriate age and volume-related information. In this
section, we complete the framework of the RT by analyzing the
transit time cdf $f(t)$ to show how complex aquifer porous volumes
can directly be quantified as a function of age, life expectancy,
and transit time.
\subsection{Characteristic groundwater volume functions.}
\begin{figure}[tbp]
\begin{center}
\includegraphics[width=\textwidth]{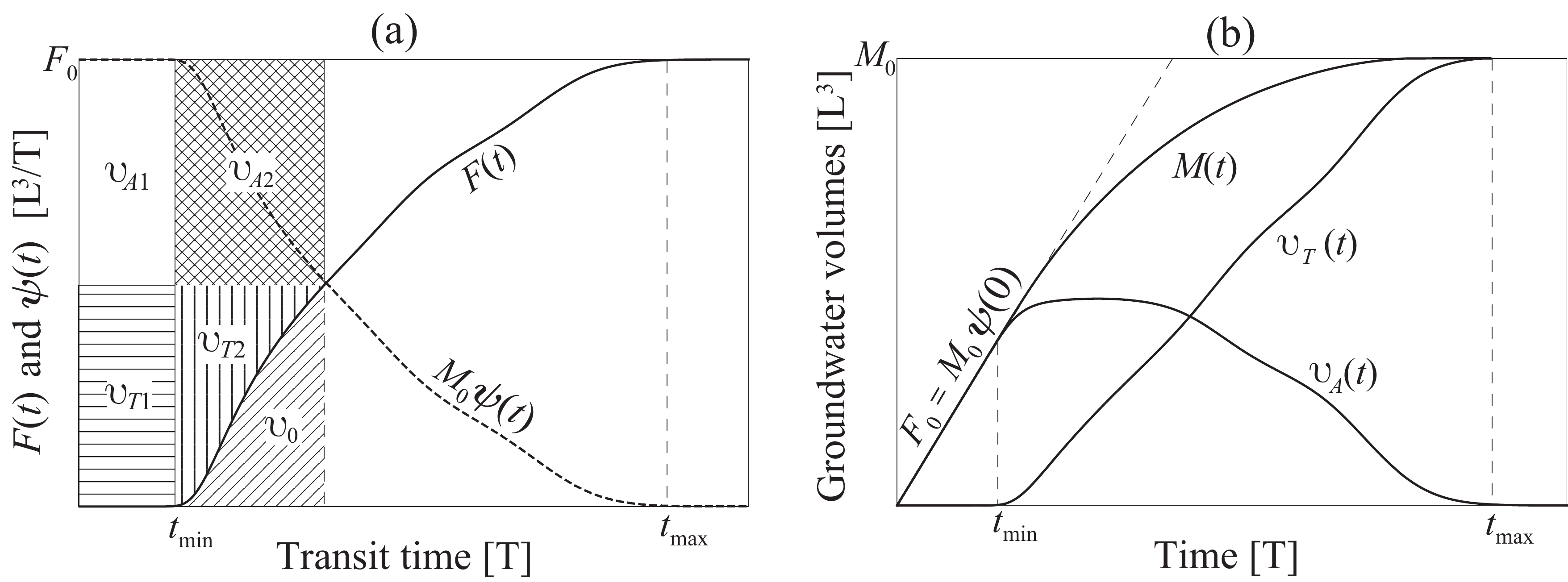}
\end{center}
\caption{Theoretical cumulated outflow function, internal age pdf
(scaled by the porous volume) and internal groundwater volume
functions: (a) Cumulated outflow function $F(t)$, with the indicated
areas $\upsilon_A$, $\upsilon_T$ and $\upsilon_0$ representing
characteristic internal groundwater volumes relative to a value of
time; (b) Aquifer porous volumes $M(t)$, $\upsilon_A(t)$ and
$\upsilon_T(t)$ as a function of age and transit
time.\label{fig:F6}}
\end{figure}

Provided the residence time distribution $\varphi(t)$ and the total
discharge $F_0$ are known, the characterization of internal
groundwater volumes can be done in a relatively straightforward
manner by a simple analysis of the function $F(t)$ as defined in
Eq.~(\ref{eq:fA}), $F(t) = F_0 \int_0^t{\varphi (u) du}$. In
Fig.~\ref{fig:F6}, the theoretical shape of the cumulated outflow
function $F(t)$ (or cumulated inflow function) is represented. The
areas $\upsilon_A$, $\upsilon_T$, and $\upsilon_0$ in
Fig.~\ref{fig:F6}a characterize different groundwater volumes that
can be defined as functions of time \textit{t} (Fig.~\ref{fig:F6}b).
The area $\upsilon_A(t)$ represents the total groundwater volume
\textit{V} in $\Omega$ of age inferior or equal to \textit{t} but
that will experience a transit time superior to \textit{t}:
\begin{equation}\label{eq:VA}
\upsilon _A (t) = V_\Omega  \left\{ {A \le t \; {\rm{and}} \; T > t}
\right\} = t [ F_0 - F(t) ] = M_0 t \psi (t)
\end{equation}

In Eq.~(\ref{eq:VA}), the pdf $\psi(t)$ is expressed given the
relationship~(\ref{eq:GRTmassbal1}) between $\psi(t)$ and $F(t)$.
When the minimum transit time $t_\textrm{min}$ is not nil, the
function $\upsilon_A(t)$ contains the volume of age inferior or
equal to time $t_\textrm{min}$, which is the area $\upsilon_{A1}(t)
= t_\textrm{min}(F_0 - F(t))$, and the volume of age inferior or
equal to \textit{t} and superior to time $t_\textrm{min}$, which is
the area $\upsilon_{A2}(t) = (t - t_\textrm{min})(F_0 - F(t))$. The
area $\upsilon_T = \upsilon_T(t)$ is the volume of groundwater that
flows through $\Omega$ with a transit time $t$ or less:
\begin{equation}\label{eq:VTT}
\upsilon _T (t) = V_\Omega  \left\{ {T \le t} \right\} = t F(t) -
\upsilon _0 (t)
\end{equation}

with the area $\upsilon_0(t)$ being the amount of exfiltrated water
having travelled from inlet to outlet during an observation time
period $t$,
\begin{equation}\label{eq:VO}
\upsilon _0 (t) = V_{\Gamma_+} \left\{ {A \le t} \right\} = \int_0^t
{F(u)du} = t F_0 - M(t)\;,
\end{equation}

and where use has been made of Eqs.~(\ref{eq:intApdf})
and~(\ref{eq:GRTmassbal1}) to express the function $M(t)$. The
quantity $tF(t)$ in Eq.~(\ref{eq:VTT}) is the total amount of
groundwater water in $\Omega$ that reaches the age \textit{t} or
less on $\Gamma_+$, plus the volume flowing out with an age
\textit{t} or less. The amount of groundwater water $\upsilon_T(t)$
is nil until the minimum transit time $t_\textrm{min}$, and reaches
the total porous volume at the maximum transit time
$t_\textrm{max}$. The function $\upsilon_T(t)$ contains the volume
of age inferior or equal to time $t_\textrm{min}$, which is the area
$\upsilon_{T1}(t) = t_\textrm{min} F(t)$ in Fig.~\ref{fig:F6}a, and
the volume of age superior to time $t_\textrm{min}$, which is the
area $\upsilon_{T2}(t) = (t - t_\textrm{min})F(t) - \upsilon_0(t)$
in Fig.~\ref{fig:F6}a. Note also that the total amount of
groundwater of age $t_\textrm{min}$ or less is given by the sum
$\upsilon_{A1} + \upsilon_{T1} = t_\textrm{min} F_0$.

Since the amount of groundwater $\upsilon_T(t)$ is the internal
volume that will leave the reservoir up to time \textit{t}, it
equals the internal transit time cumulative distribution function,
$\upsilon_T(t) = M_T(t)$, and it represents a part of the function
$M(t)$. The complementary part is the amount of groundwater water of
age \textit{t} or less and of transit time superior to \textit{t},
namely the function $\upsilon_A(t)$ defined in Eq.~(\ref{eq:VA}):
\begin{equation}\label{eq:Moft}
M(t) = V_\Omega  \left\{ {A \le t} \right\} = \upsilon_A(t) +
\upsilon_T(t)
\end{equation}

With Eq.~(\ref{eq:Moft}), one can express $\upsilon_A(t)$ as the
difference between two increasing functions that both tend to the
porous volume $M_0$ at infinity. The function $\upsilon_A(t)$ is
thus zero at the origin and at infinity. Since $\upsilon_T(t)$ is
zero until the minimum transit time $t_\textrm{min}$,
$\upsilon_A(t)$ must equal $M(t)$ between 0 and $t_\textrm{min}$.
During this time-span, which can be taken as the signature of badly
recharged and/or advection-dominated systems for significant values
of $t_\textrm{min}$, these two functions have a constant derivative
(see Fig.~\ref{fig:F6}b) equal to the steady flow rate $F_0 = M_0
\psi(0)$. The behavior of this function (number of peaks, compared
durations of increasing and decreasing parts) is instructive as to
the volumetric proportions of groundwater remaining a long time in
the system, or flowing quickly to the outlet. The time after which
the function $\upsilon_A(t)$ starts to decrease gives information on
the importance of the water volumes in the aquifer with long or
short transit times. If this time is relatively young, then the
aquifer may present a good turn-over property, and vice-versa.

Differentiating Eq.~(\ref{eq:Moft}) with respect to time, and
accounting for Eqs.~(\ref{eq:intApdf}),~(\ref{eq:intTTpdf})
and~(\ref{eq:GRT}), yields
\begin{equation}\label{eq:ERT}
\varphi(t) =  \frac{\tau_0}{t} \Psi(t)
\end{equation}
This fundamental relation includes all the features of the RT in the
most compact form. Compared to the standard rule~(\ref{eq:GRT}),
Eq.~(\ref{eq:ERT}) is a great improvement. It is simpler and may
provide the transit time distribution $\varphi(t)$ with much higher
resolution and accuracy, in relation with the fact that no
differentiation between $\varphi(t)$ and $\Psi(t)$ is required.

Note that if $F(t)$ characterizes the inlet cumulated inflow rate,
i.e. $F(t) = F_0 f_E(t)$, then each aquifer porous volume defined
above as a function of age becomes a function of life expectancy. By
analyzing the system outlet transit time cdf, the quantification of
complex groundwater volumes with respect to age, life expectancy,
and transit time, is straightforward, and provides important
practical system insights. An obvious application is the groundwater
resources protection and vulnerability assessments. The underground
storage of toxic wastes also requires the knowledge of groundwater
volumes that are potentially subject to contamination, and
assumptions on the time spent by these volumes within the reservoir
until they reach an outlet area.
\subsection{Characteristic times versus aquifer particularities.}
\label{sec:charactimes} Following Bolin and Rhode~\cite{Bolin73},
some aquifer configuration particular cases can be drawn from the
first moments of $\varphi(t)$ and $\psi(t)$.\smallskip\\
The first considerations correspond to the case for which the
average transit time and the average internal age are identical,
$\tau_\textrm{t} = \tau_\textrm{i}$. Using Eqs.~(\ref{eq:m1phi})
and~(\ref{eq:m1psi}), the condition for having $\tau_\textrm{t} =
\tau_0 = \tau_\textrm{i}$ is $\int_{0}^{\infty} {t [\varphi(t) -
\psi(t)] dt} = 0$, for which a sufficient condition is $\varphi(t) =
\psi(t) = \frac{1}{{\tau_0}}\exp \left( {-\frac{t}{{\tau_0}}}
\right)$. The exponential form of the transit time pdf is a direct
consequence of the first-order linear differential
equation~(\ref{eq:GRT}): if one of the two pdfs $\varphi(t)$ and
$\psi(t)$ has an exponential form, then the other must be identical.
This is the exponential model, often termed the well-mixed model,
which is mathematically equivalent to the unit response function of
a well-mixed reservoir. In chemical engineering, this model is used
for reactors inside which the age distribution of the elements is
uniform, i.e. there is a perfect mixing of the elements. Although
the exponential model is widely used by hydrogeologists to simulate
isotopic data, its application in aquifer systems must be handled
carefully, since it involves a large number of poorly realistic
assumptions on the aquifer structure and recharge conditions.
Eriksson~\cite{Eriksson58} interpreted the exponential distribution
of ages in groundwater as a consequence of an exponential decrease
of porosity and permeability with depth. Luther and
Haitjema~\cite{Luther98} argued that the conditions on the validity
of the exponential residence time distribution in porous media
(horizontal, un-stratified and homogeneous aquifer with respect to
porosity $\phi$, recharge rate \textit{I}, and saturated thickness
\textit{H}), can be relaxed if the parameters $\phi$, \textit{I},
and \textit{H} vary in a piecewise constant way, such that the ratio
$\phi \textit{H}/\textit{I}$ remains constant throughout the domain.
This ratio characterizes the system turnover time, which means that
each water sample taken from the reservoir must lead to a mean age
that equals the aquifer mean turnover time. In nature however, such
system configurations and conditions on mean age may hardly be
found. Etcheverry~\cite{Etcheverry01} showed that a simple linear
variation of the thickness \textit{H} significantly influences the
shape of the theoretical exponential residence time distribution.\smallskip\\
We now consider the case where $\tau_\textrm{t} > \tau_\textrm{i}$.
This case corresponds to the situation for which only few water
molecules leave the aquifer rapidly after having entered. Confined
aquifer conditions and/or very distant recharge and discharge zones
are typical settings leading to these features. For such a
configuration, the outlet transit time pdf $\varphi(t)$ is generally
small or nil for young ages, attesting the existence of a minimum
duration for travelling from inlet to outlet. The functions
$\upsilon_A(t)$ and $M(t)$ tend to remain identical until the
minimum transit time is reached. After this date, the function
$\upsilon_A(t)$ should decrease rapidly, reflecting the fact that
after the shortest travel distance from source to sink has been
covered, the outflow of older water molecules is concentrated over a
short time-span, in relation to the relative uniformity of the
travel distances within the system. A narrow triangle-shaped
function $\upsilon_A(t)$ is typical of aquifers with significant
minimum transit times and bad turnover property. Note also that from
Eq.~(\ref{eq:m1psi}) one can see that for systems with large
turnover time, or relatively low hydro-dispersive properties (i.e.
when $\sigma [\varphi]$ is small), the average age in the reservoir
may also be smaller than the turnover time, with the lowest possible
value $\tau_\textrm{i} = \frac{\tau_0}{2}$ for the piston-flow
configuration only ($\sigma [\varphi] = 0$).\smallskip\\
Finally, the case $\tau_\textrm{t} < \tau_\textrm{i}$ corresponds to
the situation for which important amounts of water molecules enter
the aquifer and flow out relatively rapidly, while sufficient
amounts of water stay long enough to increase the value of
$\tau_\textrm{i}$. For such a configuration, $\varphi(0)$ must be
bigger than $\psi(0)$, and the two curves must both decrease and
cross each other at a certain date, after which $\psi(t)$ is higher
than $\varphi(t)$. This situation may be encountered when the source
and sink zones are close to each other, or when superficial recharge
is uniformly distributed, such that the fraction of young water is
important at outlet, but when the heterogeneity of the velocity
distribution is such that long travel paths might lead to old ages
within the domain. Karstic systems are typical media where the case
$\tau_\textrm{t} < \tau_\textrm{i}$ is encountered, in relation to
the effect of the high velocities in the karstic network, which can
carry water molecules more or less independently of the surrounding
low permeability matrix. Considering Eq.~(\ref{eq:m1psi}), the case
$\tau_\textrm{t} < \tau_\textrm{i}$ can occur in systems with small
turnover time, or relatively high hydro-dispersive properties (large
$\sigma [\varphi]$).
\section{Analysis of dispersion and aquifer geometry effects on age
         distributions.}
In the following, 2-D theoretical examples are presented to
illustrate the proposed methods. Analytical and numerical solutions
are provided in order to test the sensitivity of the probability and
density functions to the advective and dispersive parameters, as
well as to the aquifer structure.
\subsection{Two-dimensional single flow system aquifer.}
\begin{figure}[tbp]
\begin{center}
\includegraphics[width=\textwidth]{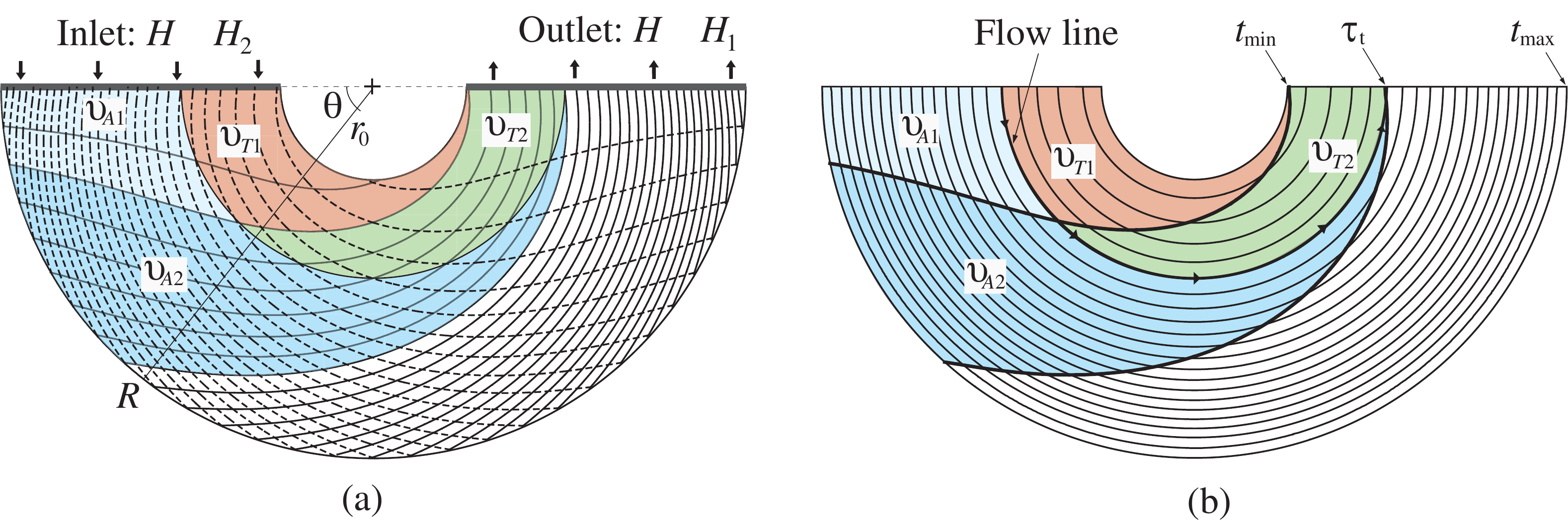}
\end{center}
\caption{Analytical solutions for mean age, mean life expectancy and
mean transit time for the 2-D vertical half-circular aquifer
(Appendix~B): (a) Mean age (solid lines) and mean life expectancy
(dashed lines) isochrones with 50~days increment; (b) Mean transit
time isochrones with 50~days increment. Parameters:
$\Delta\textit{H} = \textit{H}_2 - \textit{H}_1 = 100$~m,
$\textit{K} = 10^{-4}$~m/s, $\phi = 0.2$, inner radius $r_0 =
250$~m, outer radius $R = 1000$~m, minimum transit time
$t_\textrm{min} = 143$~days, maximum transit time $t_\textrm{max} =
2285$~days, turnover time $\tau_0 = \tau_\textrm{t} =
772.5$~days.\label{fig:F8}}
\end{figure}

\begin{figure}[tbp]
\begin{center}
\includegraphics[width=\textwidth]{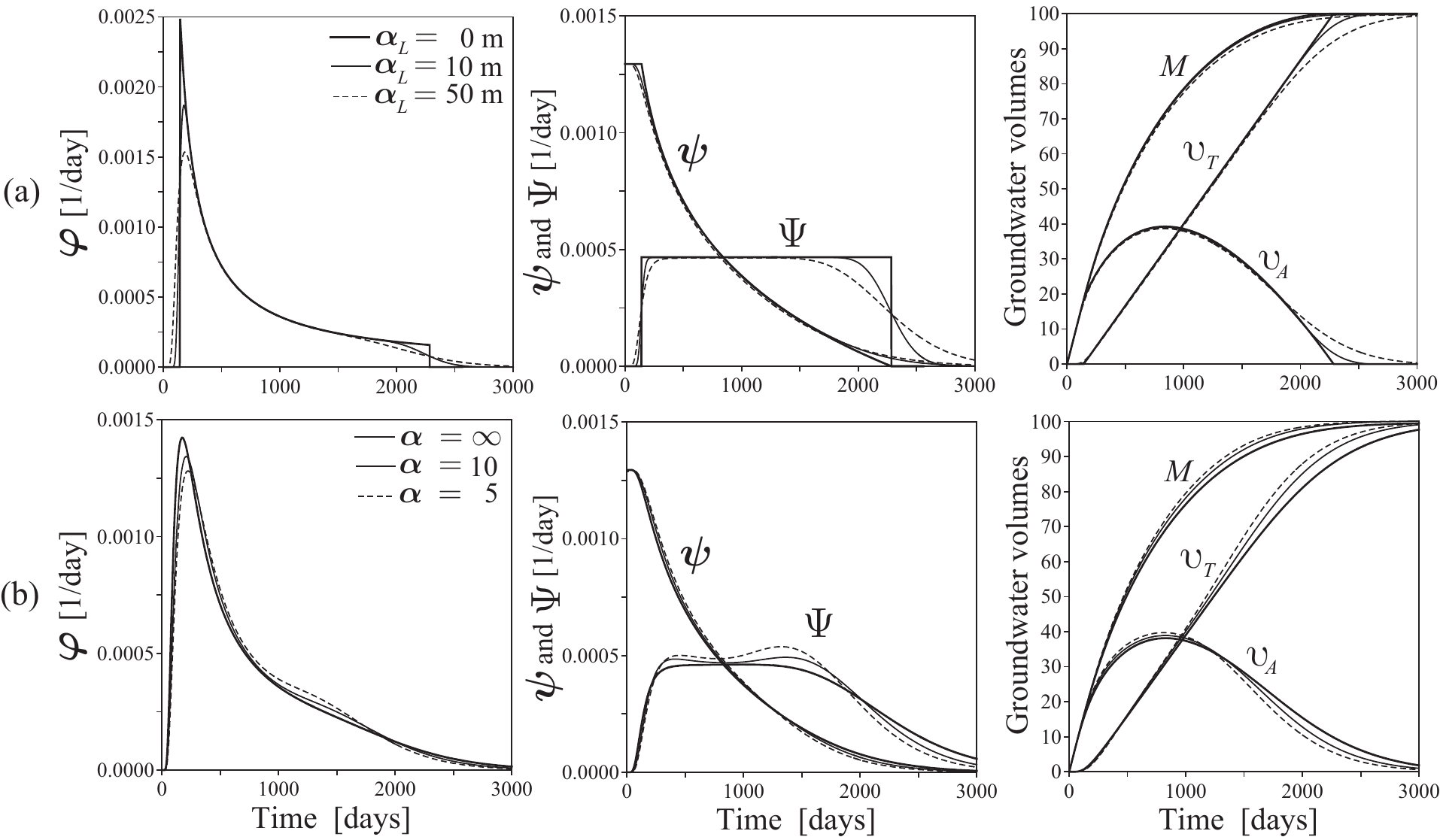}
\end{center}
\caption{RT solutions for the 2-D vertical half-circular aquifer:
(a) Analytical solutions (see Appendix~B) for the pdfs $\varphi(t)$,
$\psi(t)$ and $\Psi(t)$, and for the internal groundwater volume
functions (in \% of $M_0$) $M(t)$, $\upsilon_A(t)$ and
$\upsilon_T(t)$, as a function of the $\alpha_L$ coefficient
($\alpha_T = D_\textrm{m} = 0$); (b) Numerical solutions as a
function of the ratio $\alpha = \alpha_L/\alpha_T$, with $\alpha_L =
50$~m.\label{fig:F9}}
\end{figure}

A 2-D half-circular crown-shaped reservoir is used to simulate  a
single-flow system aquifer, homogeneous with respect to porosity
$\phi$ and hydraulic conductivity \textit{K}. A positive head
difference is applied between the recharge and discharge areas
(Fig.~\ref{fig:F8}a). This geometry is well suited for the
derivation of analytical solutions~\cite{Etcheverry00}. Due to
homogeneity and symmetry, the flow lines remain parallel to each
other, flow being one-dimensional along the flow line coordinate.
The spatial distributions of mean age and mean life expectancy can
thus be solved analytically (see Appendix B), and reveal to be
symmetric (Fig.~\ref{fig:F8}a), yielding a co-centrical distribution
of mean transit time. The porous volumes $\upsilon_A$ ($=
\upsilon_{A1} + \upsilon_{A2}$) and $\upsilon_T$ ($= \upsilon_{T1} +
\upsilon_{T2}$), quantified using the cumulative outflow function
$F(t)$, can easily be identified in Fig.~\ref{fig:F8} by using the
mean age and the mean transit time isochrones. Note that the special
case of this theoretical aquifer allows a good representation of the
groundwater internal volumes with the mean values of the global
distributions. However, the mean of the age, life expectancy, and
transit time distributions would generally give a rough
representation only, when e.g. significant dispersive processes take
place.
Etcheverry and Perrochet~\cite{Etcheverry00} analyzed the pure
advective case and found that $\varphi(t)$ is proportional to $1/t$,
and that $\psi(t)$ is proportional to $\ln(1/t)$ between the minimum
and the maximum transit time (see Fig.~\ref{fig:F9}a). The function
$\Psi(t)$ is a constant from the minimum to the maximum transit
time, which indicates that the aquifer volumes ranging between a
unit increase of transit time remain constant (the aquifer volumes
between two iso-contours in Fig.~\ref{fig:F8}b are all the same).
The case $\alpha_T = 0$ in Fig.~\ref{fig:F9}a was analyzed by means
of analytical solutions (Appendix~B). For this system, the pdfs
$\varphi(t)$, $\psi(t)$ and $\Psi(t)$ show generally moderate
fluctuations with respect to longitudinal dispersion. The main
effect of the $\alpha_L$ coefficient can be seen in the increase of
the spreading of the probability distributions. The outlet transit
time pdf shows younger arrival times when $\alpha_L$ increases, but
also a longer tail. The functions $\psi(t)$ and $\Psi(t)$ are
affected the same way by longitudinal dispersion. The average
internal age $\tau_\textrm{i}$ and the average internal transit time
$\tau_\textrm{it}$ are dispersion dependent; they increase linearly
with $\alpha_L$, indicating that dispersion leads to a higher
average age of the system. Note that the second temporal moments of
the internal age and transit time distributions increase
proportionally with the square of $\alpha_L$, while the variance of
$\varphi(t)$ increases linearly with $\alpha_L$.
Contrasted behavior appears when the coupled effect of longitudinal
and lateral dispersivity is taken into account (Fig.~\ref{fig:F9}b).
The case $\alpha_T \neq 0$ was analyzed using numerical solutions,
by increasing the ratio $\alpha = \alpha_L/\alpha_T$. The tailing
effect decreases from a starting situation ($\alpha = \infty$ in
Fig.~\ref{fig:F9}b) with increasing values of lateral dispersivity
(decrease of $\alpha$ ratio in Fig.~\ref{fig:F9}b). Lateral
dispersivity induces the mixing of ages, old water molecules moving
laterally between the flow lines and replacing younger molecules,
and vice-versa. The internal groundwater volume functions also
attest of the above-mentioned dispersion effects. For example, the
function $\upsilon_A(t)$ shows long tails for large values of
$\alpha_L$, reinforcing the observed behavior at outlet of the
transit time pdf for old ages. The internal increase of water
volumes that require long transit times to exit the aquifer due to
an increase of $\alpha_L$ has of course its consequence at outlet
with older arrival times. As with the pdfs $\varphi(t)$, $\psi(t)$
and $\Psi(t)$, this tailing effect decreases when lateral dispersion
is added.
\subsection{Vertical multi-layer aquifer.}
\begin{figure}[tbp]
\includegraphics[width=\textwidth]{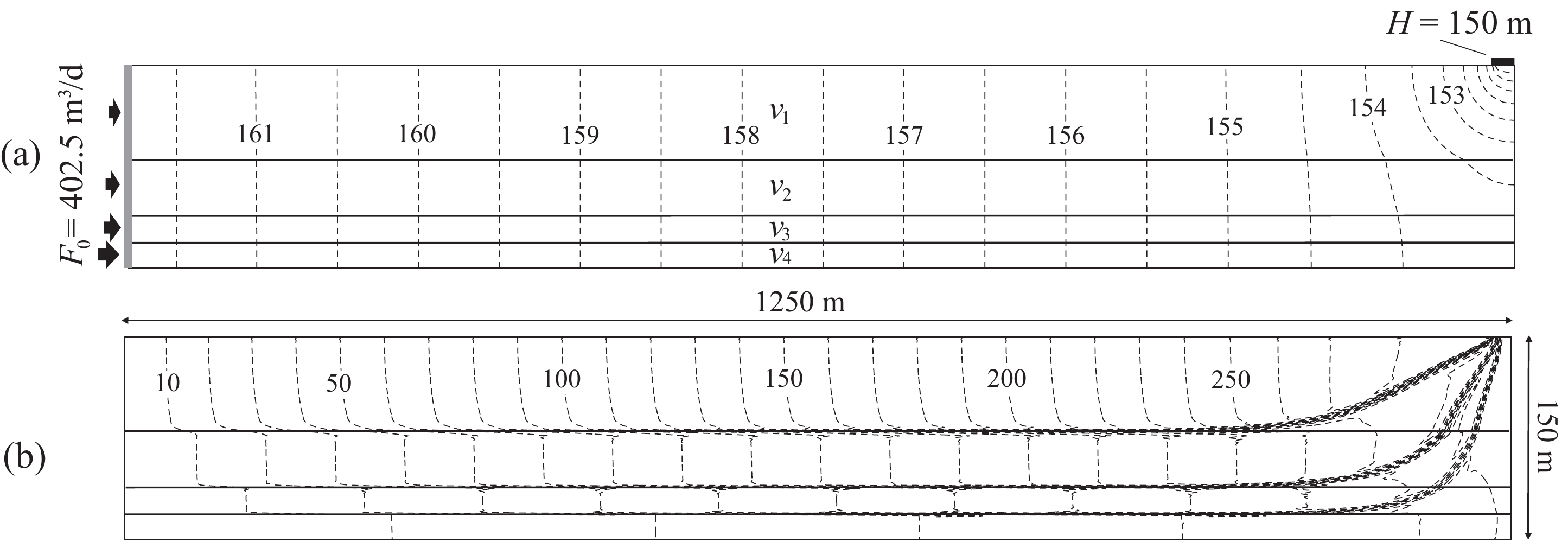}
\caption{2-D vertical multi-layered aquifer: (a) Boundary conditions
and Laplacian flow field in meters; (b) Mean age isochrones in days,
illustrating the pore velocity contrast effect on age transport
($\alpha_L = \Delta x = \Delta z = 2.5~\textrm{m}$, $\alpha_T = 0$,
$D_\textrm{m} = 0$). Average pore velocities (in m/day): $v_1 =
4.0$, $v_2 = 6.5$, $v_3 = 10.5$, $v_4 = 23.5$.\label{fig:F10}}
\end{figure}

\begin{figure}[tbp]
\begin{center}
\includegraphics[width=\textwidth]{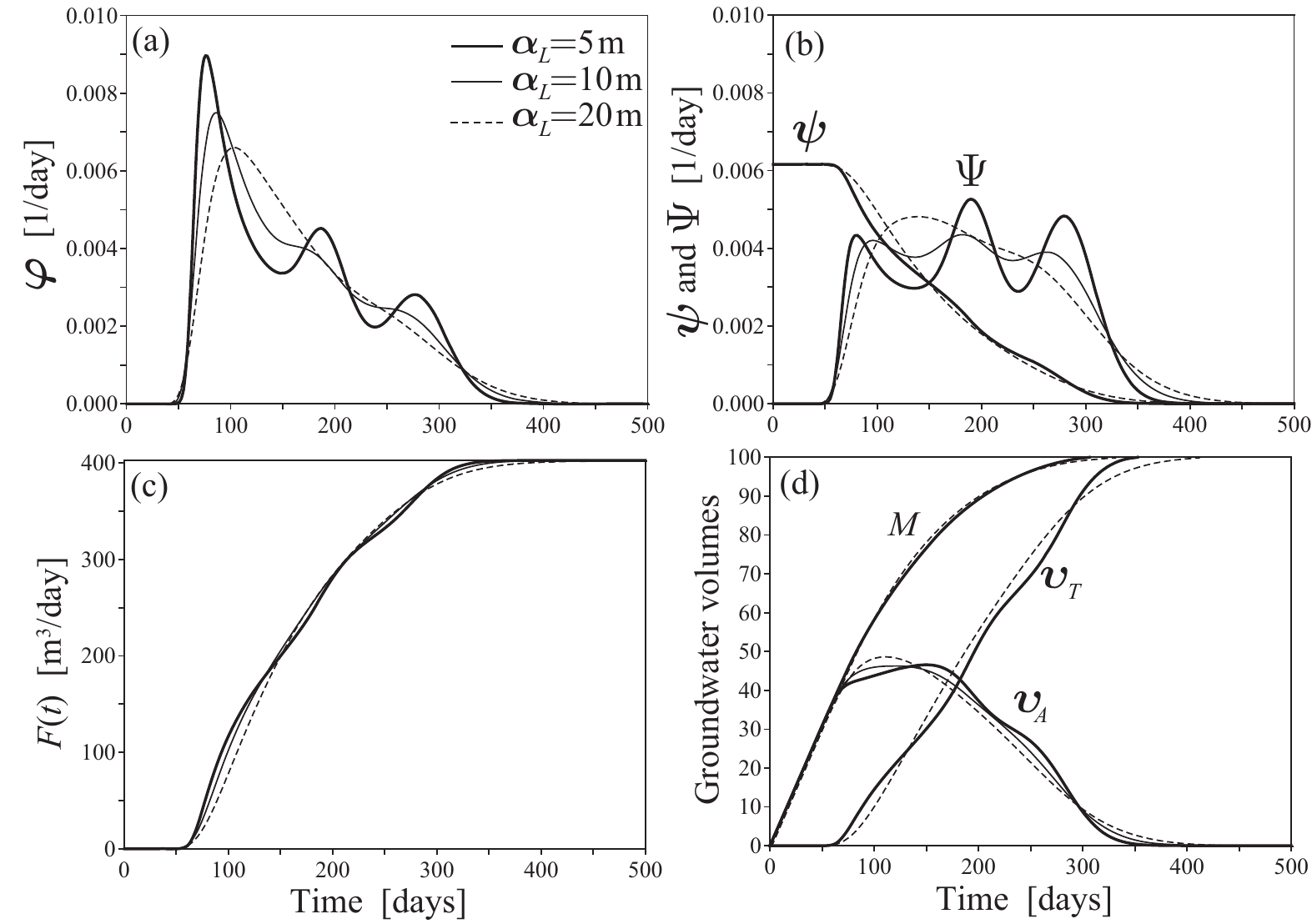}
\end{center}
\caption{Calculated pdfs, cdfs and groundwater volumes for the 2-D
vertical multi-layered aquifer as a function of dispersion, for the
ratio $\alpha = \alpha_L/\alpha_T = 10$: (a) Outlet transit time
pdf; (b) Internal age pdf and internal transit time pdf; (c) Outlet
cumulated outflow function; (d) Groundwater volumes versus time (in
\% of $M_0$).\label{fig:F11}}
\end{figure}

\begin{figure}[tbp]
\begin{center}
\includegraphics[width=\textwidth]{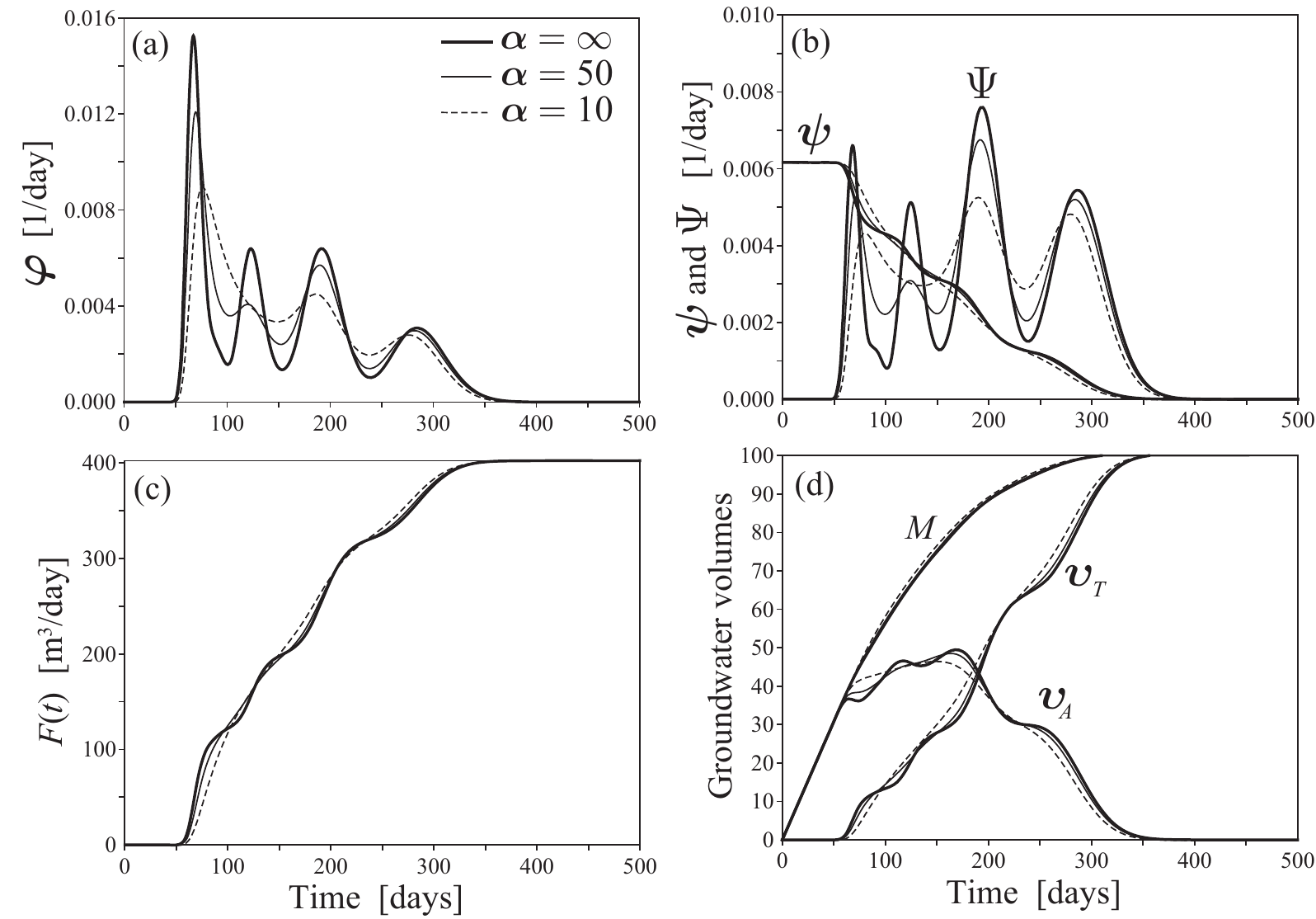}
\end{center}
\caption{Calculated pdfs, cdfs and groundwater volumes for the 2-D
vertical multi-layered aquifer as a function of the ratio $\alpha =
\alpha_L/\alpha_T$: (a) Outlet transit time pdf; (b) Internal age
pdf and internal transit time pdf; (c) Outlet cumulated outflow
function; (d) Groundwater volumes versus time (in \% of
$M_0$).\label{fig:F12}}
\end{figure}

In this second theoretical example, we consider a four-layered
vertical aquifer system, as illustrated in Fig.~\ref{fig:F10}a. From
the top to the bottom of the model, the layers have decreasing
thicknesses and increasing pore velocities. The domain is
discretized into 30'000 homogeneous bilinear quadrangles. The total
porous volume $M_0$ is 65312.5~$\textrm{m}^{3}$. A constant input
flow rate enters the system along the left limit, using imposed
fluxes of varying intensity proportional to the different layers
hydraulic conductivities. The system turnover time is $\tau_0 =
162.267$~days. The outlet is simulated at the top of the right side
by a constant hydraulic head along a relative small zone of 15~m.
The permeability-porosity couples have been set in a way that the
pore velocity contrasts involve specific ages within each layer
(velocities $v_i$ in Fig.~\ref{fig:F10}a), creating specific ages
within each layer, as illustrated in Fig.~\ref{fig:F10}b. The fact
that the layers thickness diminishes with depth, while the influx
intensity increases, is meant to create arrival time peaks at outlet
of comparable orders of magnitude.
The temporal moments $\tau_\textrm{i}$ and $\tau_\textrm{it}$ are
good indicators of the dynamics of the global system. Because they
are volume-averaged quantities, their magnitude will depend on the
water quantities of contrasted age and transit time, which are
directly related to the flow and transport dynamics. For this
example, $\tau_\textrm{i}$ and $\tau_\textrm{it}$ show only small
variations with respect to dispersion (Table~\ref{tab:T1}), because
even if velocities between layers are contrasted, flow in the system
is generally rapid. Longitudinal dispersion has the effect of making
the system older, by creating long tails that can be observed at the
reservoir outlet, but also on the internal age and internal transit
time distributions (Figs.~\ref{fig:F11}a and~\ref{fig:F11}b).
Spreading in the direction of velocity is of course the main cause
for this. Lateral dispersion presents the property of homogenizing
the ages by mixing water molecules of different ages. For a given
level of longitudinal dispersion, increasing the ratio $\alpha =
\alpha_L/\alpha_T$ (Fig.~\ref{fig:F12}) makes the system younger
(tailing is lowered), with diminishing values of the mean internal
age $\tau_\textrm{i}$ and the mean internal transit time
$\tau_\textrm{it}$ (Table~\ref{tab:T1}). Longitudinal dispersion
induces spreading of solute particles, and thus variability around
the plume center of gravity. When lateral dispersion is added, the
mixing of water molecules between flow lines homogenizes the ages.
When longitudinal spreading is important, the exchange surface along
the plume body is extended, and mixing can then be expected to be of
high magnitude.
The internal transit time pdf $\Psi(t)$ is of particular interest
for the characterization of the internal organization of the flow
dynamics. Its shape and particularly the number of modes it shows
has a direct consequence on the shape of the outlet transit time
pdf. If we take the example of the case $\alpha = \infty~(\alpha_T =
0)$ in Fig.~\ref{fig:F12}, the function $\Psi(t)$ exhibits as many
peaks as the outlet transit time pdf $\varphi(t)$. However, the
magnitude of these peaks are different for the two functions, and
discern the transit time frequencies at outlet from those of the
system pore volume. The peak of maximum intensity is the first one
for $\varphi(t)$, and the third one for $\Psi(t)$. If one looks at
the time value corresponding to the first peak of $\varphi(t)$, say
$t_1$, then inside the domain the density of probability that the
water molecules have a transit time equal to $t_1$ is smaller than
the same density of probability at the reservoir outlet. This is due
to the fact that $\Psi(t)$ deals with volume-average probabilities
while $\varphi(t)$ deals with flux-average probabilities. The four
consecutive peaks of the outlet transit time pdf correspond to the
four families of water molecules, which transit within the four
layers of the system, even though some parts of the curve attest to
some mixing effects (the density of probability is not necessarily
zero between two peaks).
\begin{table}
\caption{Dispersive parameters related to Figs.~\ref{fig:F11}
and~\ref{fig:F12}, and calculated pdf statistics (units in meters
and days).} \centerline{
\begin{tabular}{|c|ccc|ccc|}
\hline \textbf{Parameters} & \textbf{Fig.~\ref{fig:F11}} & $$ & $$ &
\textbf{Fig.~\ref{fig:F12}} & $$ & $$ \\
\hline
$\alpha = \alpha_L/\alpha_T$ & $10$ & $10$ & $10$ & $\infty$ & $50$ & $10$ \\
$\alpha_L$ & $5.0$ & $10.0$ & $20.0$ & $5.0$ &
$5.0$ & $5.0$ \\
$\alpha_T$ & $0.5$ & $1.0$ & $2.0$ & $0.0$ & $0.1$
& $0.5$ \\
\hline
\textbf{Statistics} & $$ & $$ & $$ & $$ & $$ & $$ \\
\hline $\sigma[\varphi]$ & $75.710$ & $75.955$ & $79.871$ &
$81.766$ & $79.786$ & $75.940$ \\
$\tau_\textrm{i}$ & $98.902$ & $98.915$ & $101.010$ &
$102.156$ & $100.905$ & $98.916$ \\
\hline
\end{tabular}}
\label{tab:T1}
\end{table}
This simple theoretical example shows the complexity that is to be
expected for the nature of the age and transit time distributions.
It underlines how the significance of average age can lead to
erroneous interpretations, regarding the system dynamics (e.g.
inferences on average velocity) or hydrogeological problems related
to risk and vulnerability assessment. Whenever the outlet transit
time pdf is multi-modal, then the average transit time can
definitely not be used as a reliable quantity representative of the
distribution since, as illustrated here, mean ages at outlet are
mostly related to the lowest densities of probability (see
Figs.~\ref{fig:F11} and~\ref{fig:F12},~Table~\ref{tab:T1}). Once
more this consideration points out to the interpretation of average
age measurements, but also mean age simulations, which should be
done very carefully. Generally, attention is given to hydrodynamic
dispersion, heterogeneity, and long distance travelling induced
mixing~\cite{Goode96,Varni98,Weissmann02}, which are factors that
represent a source of uncertainty for the age dating methods and
environmental tracer data interpretations. Here one can see that the
geological, the structural and the hydraulic boundary configurations
can by themselves be responsible of unrepresentative mean ages, even
in advection-dominated transport regimes. Even if the entire
groundwater age distribution cannot be measured in the field, mean
age dates can be of great help for model calibration purposes.
However, mean ages should a priori not represent an absolute
simulation answer, and the knowledge of the entire age distribution
should be of prime interest in most cases.
\section{Summary and conclusions.}
(1) The probability distributional evolution of groundwater age and
life expectancy has been simulated by forward and backward transient
advection-dispersion type equations, according to proper boundary
conditions. For a given position in space, the age and the life
expectancy pdfs are complementary distributions. Straight
application of the principle of superposition to these distributions
results in an integrated distribution of transit times, which
characterizes the probability of having a given transit time from
the recharge zone to the discharge zone, together with the
quantities associated to that transit time. The transit time pdf is
simply the convolution integral of the age pdf and the life
expectancy pdf, and tells about the entire history of water
molecules since recharge until discharge. Age, life expectancy, and
transit time pdfs provide different kind of information, and each of
them can reveal to be more advantageous than the other one depending
on the hydrogeological insights to be provided. For instance, life
expectancy and transit time distributions are well suited for
vulnerability assessment problems (e.g. well-head protection, or
underground storage of toxic waste), and they allow the mapping of
different regions within a recharge zone, in terms of residence time
in the aquifer and associated properties.

(2) The results of Eriksson~\cite{Eriksson71} have been recovered by
manipulating the advection-dispersion equation, extending thus the
RT to systems with significant dispersive components. In the
classical RT neither advection nor dispersion was considered. We
have demonstrated here that the RT is still valid in systems with
spatially varying velocity fields and non-negligible
dispersive/diffusive effects. The RT is a simple one-dimensional
formulation, time being the only dependent variable. The outlet
transit time distribution is derived from the internal age
distribution, and therefore has a much more refined resolution. In
so doing, mixing of converging flow patterns near the outlet is
ignored, and the maximum transit time is never smaller than the
maximum age in the reservoir. The RT formulation also applies to the
internal distribution of life expectancy, and can be used to
evaluate recharge zones life expectancy pdfs. It has also been shown
that the RT can be expressed in a more compact form, which relates
the outlet transit time distribution to its internal counterpart,
the internal distribution of transit times from inlet to outlet.

(3) From the reservoir characteristic distributions, fundamental
additional transient information on water volumes and water fluxes
can be gained. From the outlet transit time cdf, specific
groundwater porous volumes can directly be identified and quantified
with respect to age, life expectancy and transit time. These
functions can be very useful for aquifer characterization and
intrinsic vulnerability assessments. They can be used to easily
evaluate the magnitudes of young and old groundwater volumes in the
aquifer.

(4) Using analytical and numerical solutions for theoretical aquifer
configurations, some effects of macro-dispersion on simulated age
and transit time distributions could be underlined. For instance,
longitudinal dispersion can have a significant aging effect, while
lateral dispersion rejuvenates the system through transverse mixing.
Like in previous studies, it has been shown that the average age
resulting from dating-methods, or direct simulations, can lead to
erroneous interpretations. It is a well-known fact that dispersion
can induce a high variability of the age distribution around the
average. Here we have shown that not only the mixing processes could
make the average age insignificant, but also the geological
structure and the geometry of the flow patterns.

(5) The proposed methodology can equivalently be implemented in
one--, two-- and three--dimensions, and has considerable technical
and numerical advantages, which may be pivotal in handling very
large systems. In fact, when the outlet transit time pdf is defined
by integrating all hydro-dispersive properties over the entire flow
field, the level of refinement required by a stable transport model
is generally sufficient. The RT ensures that the minimum and the
maximum age in the reservoir are captured at the outlet, which is
hardly the case with traditional methods, mainly because of the
mixing of converging fluxes near the outlet. In the present work,
the RT has been combined with ADEs, and the equations were solved
using the LTG technique. However, no restriction appears for the use
of the methodology if age and life expectancy pdfs are calculated by
other transport models, such as random-walk simulators.

(6) The presented models have been developed for the global aquifer
system, regardless the number of individual inlet and outlet zones.
At this stage, the RT has been rendered applicable to the whole
system. In a subsequent article (this issue), we generalize the RT
to any observation zone, and to systems with several inlets and
outlets.

\ack{The authors would like to acknowledge the Swiss Research
National Fund for financially supporting this research under Grant
no. 2100-064927. The authors also thank Professor H.-J. Diersch and
one anonymous reviewer for constructive criticism of this
manuscript.}
\renewcommand{\theequation}{A.\arabic{equation}}
\setcounter{equation}{0}
\section*{Appendix A: Reservoir Theory for a 1-D semi-infinite
         domain.}
\label{ap:A} The age resident pdf is found by solving the
one-dimensional form of the age pdf ADE~(\ref{eq:A ADE}) with the
Cauchy type condition $vg_A (0,t) - D{{\partial g_A (0,t)}
\mathord{\left/{\vphantom {{\partial g_A (0,t)} {\partial x}}}
\right.\kern-\nulldelimiterspace} {\partial x}} = v \delta (t)$, and
with a zero concentration gradient at a point at infinity. The
corresponding age flux pdf is the flux concentration $g_A^f$, which
is deduced using the one-dimensional form of Eq.~(\ref{eq:A flux
conc}), $g_A^f = g_A - \frac{D}{v}\frac{\partial g_A}{\partial x}$.
These solutions are e.g. given in Jury and Roth~\cite{Jury90}. Using
the dimensionless variables
\[ \textrm{X} = \frac{x}{L} \quad\quad
\textrm{T} = \frac{v}{L}t \quad\quad \textrm{Pe} = \frac{{vL}}{D} \]
where \textit{L} is a characteristic length defining the supposed
outlet position, and where $\textrm{Pe}$ is the P\'{e}clet number,
the age resident and flux pdfs read
\begin{equation}\label{eq:a1}
g_A (\textrm{X},\textrm{T}) = \frac{{\sqrt {{\textrm{Pe}}} }}{{\sqrt
{\pi \textrm{T}} }}\exp \left( { - \frac{{{\textrm{Pe}}(\textrm{X} -
\textrm{T})^2 }}{{4\textrm{T}}}} \right) -
\frac{{{\textrm{Pe}}}}{{\rm{2}}}\exp (\textrm{X} {\textrm{Pe}})
\erfc \left( {\frac{{\sqrt {{\textrm{Pe}}} (\textrm{X} +
\textrm{T})}}{{2\sqrt \textrm{T} }}} \right)
\end{equation}
\begin{equation}\label{eq:a2}
g_A^f (\textrm{X},\textrm{T}) = \frac{{\textrm{X}\sqrt
{{\textrm{Pe}}} }}{{2\sqrt \pi \textrm{T}^{\frac{3}{2}} }}\exp
\left( { - \frac{{{\textrm{Pe}}(\textrm{X} - \textrm{T})^2
}}{{4\textrm{T}}}} \right)
\end{equation}
The life expectancy resident and flux pdfs can be deduced
from~(\ref{eq:a1}) and~(\ref{eq:a2}) by substitution of $\textrm{X}$
by $1 - \textrm{X}$ and $\textrm{X}$ by 1, respectively:
\begin{eqnarray}\label{eq:a3}
g_E (\textrm{X},\textrm{T}) & = & \frac{{\sqrt {{\textrm{Pe}}}
}}{{\sqrt {\pi \textrm{T}} }}\exp \left( { - \frac{{{\textrm{Pe}}(1
- \textrm{X} - \textrm{T})^2 }}{{4\textrm{T}}}} \right)
\nonumber\\
& - & \frac{{{\textrm{Pe}}}}{{\rm{2}}}\exp [(1 - \textrm{X})
{\textrm{Pe}}] \erfc \left( {\frac{{\sqrt {{\textrm{Pe}}} (1 -
\textrm{X} + \textrm{T})}}{{2\sqrt \textrm{T} }}} \right)
\end{eqnarray}
\begin{equation}\label{eq:a4}
g_E^f (\textrm{X},\textrm{T}) = \frac{{(1 - \textrm{X})\sqrt
{{\textrm{Pe}}} }}{{2\sqrt \pi \textrm{T}^{\frac{3}{2}} }}\exp
\left( { - \frac{{{\textrm{Pe}}(1 - \textrm{X} - \textrm{T})^2
}}{{4\textrm{T}}}} \right)
\end{equation}
The transit time flux pdf is calculated by straight application in
the time domain of the convolution integral in Eq.~(\ref{eq:TT
PDF}):
\begin{eqnarray}\label{eq:a5}
g_T^f (\textrm{X},\textrm{T}) & = & \int_{0}^{\textrm{T}} { g_A^f
(\textrm{X},\textrm{T}^{'}) g_E^f
(\textrm{X},\textrm{T}-\textrm{T}^{'}) d \textrm{T}^{'} } =
\frac{{\sqrt {{\textrm{Pe}}} }}{{2\sqrt \pi \textrm{T}^{\frac{3}{2}}
}} \exp \left( { - \frac{{{\textrm{Pe}}(1 -
\textrm{T})^2}}{{4\textrm{T}}}} \right)
\nonumber\\
& = & g_A^f (1,\textrm{T}) = g_E^f (0,\textrm{T})
\end{eqnarray}
The transit time flux pdf in Eq.~(\ref{eq:a5}) could have been
deduced by substitution of $\textrm{X}$ by 1 in Eq.~(\ref{eq:a2}),
or by substitution of $\textrm{X}$ by 0 in Eq.~(\ref{eq:a4}). The
transit time resident pdf is found by convoluting~(\ref{eq:a1})
and~(\ref{eq:a3}) in the Laplace domain:
\begin{equation}\label{eq:a6}
\hat g_T (\textrm{X},s) = \hat g_A (\textrm{X},s) \hat g_E
(\textrm{X},s) = \frac{{\rm{4}}}{{\left( {1 + \gamma } \right)^2
}}\exp \left( {\frac{{{\textrm{Pe}}}}{2}\left[ {1 - \gamma }
\right]} \right)
\end{equation}
where $s$ is the Laplace variable, $\hat g$ is
$\textit{s}$-transformed state of the function $g$, and with $\gamma
= \sqrt{(1 + 4s/\textrm{Pe})}$. The inversion of Eq.~(\ref{eq:a6})
yields
\begin{eqnarray}\label{eq:a7}
g_T (\textrm{X},\textrm{T}) & = & g_T (\textrm{T}) \nonumber\\
& = & {\textrm{Pe}}\left( {1 + \frac{{{\textrm{Pe}}(1 +
\textrm{T})}} {{\rm{2}}}} \right)\exp ({\textrm{Pe)}} \erfc \left(
{\frac{{{\textrm{Pe(1}} + \textrm{T})}}{{2\sqrt
{{\textrm{Pe}}\textrm{T}}}}} \right) \nonumber\\
& - & \frac{{{\textrm{Pe}}^{\rm{2}} \textrm{T}}}{{\sqrt
{\pi{\textrm{Pe}}\textrm{T}} }}\exp \left( { -
\frac{{{\textrm{Pe(1}} - \textrm{T})^2 }}{{4\textrm{T}}}} \right)
\end{eqnarray}
where use has been made of the shifting theorem
\[ \mathcal{L}^{- 1} \left\{ {f(a s + b)} \right\}
= \frac{{e^{- \frac{{b t}}{a}} }}{a}f\left( {\frac{{t}}{a}}
\right),\] and of the following Laplace transform:
\begin{equation*}
\begin{split}
\mathcal{L}^{- 1} \left\{ {\frac{{\exp ( - c\sqrt s )}}{{(b + \sqrt
s )^2 }}} \right\} & = (1 + bc + 2b^2 t)\exp(b^2 t + bc) \erfc
\left( {\frac{c}{{2\sqrt t}} + b\sqrt t } \right) \\
& - \frac{2bt}{\sqrt{\pi t}} \exp \left( { - \frac{{c^2 }}{{4t}}}
\right)
\end{split}
\end{equation*}
The internal age, internal life expectancy and internal transit time
pdfs are obtained from Eqs.~(\ref{eq:intApdf}),~(\ref{eq:intLEpdf})
and~(\ref{eq:intTTpdf}):
\begin{eqnarray}\label{eq:a8}
\psi ({\textrm T}) & = & \int_{0}^{1} {g_A ({\textrm X},{\textrm T})
d{\textrm X}} = \int_{0}^{1} {g_E ({\textrm X},{\textrm T})
d{\textrm X}} \nonumber\\
& = & \frac{1}{2}\left( \erfc \left( {\frac{{\sqrt {{\textrm{Pe}}}
({\textrm T} - 1)}}{{2\sqrt {\textrm T} }}} \right) - \exp
({\textrm{Pe}})\erfc \left( {\frac{{\sqrt {{\textrm{Pe}}} ({\textrm
T} + 1)}}{{2\sqrt {\textrm T} }}} \right) \right)
\end{eqnarray}
\begin{equation}\label{eq:a9}
\Psi ({\textrm T}) = \int_{0}^{1} {g_T ({\textrm X},{\textrm
T})d{\textrm X}} = g_T ({\textrm T})
\end{equation}
In 1-D, the reservoir theory is equivalent to the mass conservation
relation between flux and resident concentrations given by Jury and
Roth~\cite{Jury90}:
\begin{eqnarray}\label{eq:a10}
\varphi (\textrm{T}) & = & - \frac{{\partial \psi
(\textrm{T})}}{{\partial \textrm{T}}} = - \frac{\partial }{{\partial
\textrm{T}}} \int_0^1 {g(\textrm{X},\textrm{T})d\textrm{X}} =
\frac{{\sqrt {{\textrm{Pe}}} }}{{2\sqrt \pi \textrm{T}^{\frac{3}{2}}
}}\exp \left( { - \frac{{{\textrm{Pe}}(1 -
\textrm{T})^2 }}{{4\textrm{T}}}} \right) \nonumber\\
& = & g_A^f (1,\textrm{T}) = g_E^f (0,\textrm{T})
\end{eqnarray}
The dimensionless outlet transit time pdf is simply the age flux
concentration at $\textrm{X} = 1$, or equivalently the life
expectancy flux concentration at $\textrm{X} = 0$. The mean of the
pdfs $\varphi(\textrm{T})$, $\psi(\textrm{T})$, and
$\Psi(\textrm{T})$ read
\begin{equation}\label{eq:a11}
\tau _{\textrm{t}} = \mu _1 [\varphi ] = 1 \quad\quad  \tau
_{\textrm{i}} = \mu _1 [\psi ] = \frac{1}{2} +
\frac{1}{{{\textrm{Pe}}}} \quad\quad \tau _{{\textrm{it}}}  = \mu _1
[\Psi ] = 1 + \frac{2}{{{\textrm{Pe}}}} = 2\tau _{\textrm{i}}
\end{equation}
and the spreading of these pdfs is measured by their variance:
\begin{equation}\label{eq:a12}
\sigma ^2 [\varphi ] = \frac{2}{{{\textrm{Pe}}}} \quad\quad \sigma
^2 [\psi ] = \frac{{({\textrm{Pe}}{\rm{ + }}{\rm{6}})^2
}}{{{\rm{12}}{\textrm{Pe}}^2 }} \quad\quad \sigma ^2 [\Psi ] =
\frac{{2{\textrm{Pe}} {\rm{ + }} {\rm{6}}}}{{{\textrm{Pe}}^2 }}
\end{equation}
\renewcommand{\theequation}{B.\arabic{equation}}
\setcounter{equation}{0}
\section*{Appendix B: Reservoir Theory in a single flow system.}
\label{ap:B} Consider the crown-shaped aquifer geometry in
Fig.~\ref{fig:F8}. Since the flow lines remain parallel to each
other in the entire domain, we assume a 1-D transport process at
each point $x = \theta r$ on the flow line coordinate. We follow the
same procedure than in Appendix~A, given that the porous volume is
$M_0 = \phi \pi (R^2 - r_{0}^2)/2$, the steady flow rate $F_0 = K
\Delta H \ln(R/r_0)/\pi$, the dispersion coefficient $D(r) =
\alpha_L v(r)$, and the pore velocity $v(r) = q(r)/\phi = K \Delta H
/\phi \pi r$. The age resident and flux pdfs are deduced from
Eqs.~(\ref{eq:a1}) and~(\ref{eq:a2}) by substituting $x$ by $\theta
r$, and the life expectancy resident and flux pdfs are deduced by
replacing $x$ by $(\pi - \theta) r$. The transit time flux pdf is
finally obtained by replacing $x$ by $\pi r$ in Eq.~(\ref{eq:a2}).
The average age, life expectancy and transit time are given by the
first moment of the corresponding flux pdfs:
\begin{equation}\label{eq:b1}
\langle A\rangle (\theta ,r) = \frac{{\theta \phi \pi r^2
}}{{K\Delta H}}
\end{equation}
\begin{equation}\label{eq:b2}
\langle E\rangle (\theta ,r) = \frac{{(\pi  - \theta ) \phi \pi r^2
}}{{K\Delta H}}
\end{equation}
\begin{equation}\label{eq:b3}
\langle T\rangle (r) = \langle A\rangle (\theta ,r) + \langle
E\rangle (\theta ,r) = \frac{{ \phi \pi^2 r^2 }}{{K\Delta H}}
\end{equation}
Since on the outlet line $(\theta = \pi)$ velocity points in the
direction of the outward unit vector, the discharge zone transit
time pdf $\varphi(t)$ can be evaluated by averaging the age flux
pdfs on the outlet line, or equivalently by averaging the life
expectancy flux pdfs on the inlet line $(\theta = 0)$. The internal
age (or life expectancy) pdf $\psi(t)$, and the internal transit
time pdf $\Psi(t)$ are calculated following the domain integration
in Eqs.~(\ref{eq:intApdf}),~(\ref{eq:intLEpdf})
and~(\ref{eq:intTTpdf}), by integrating between the inner radius
$r_0$ and the outer radius $R$, and by integrating between 0 and
$\pi$ (see Fig.~\ref{fig:F8}). The average transit time
$\tau_\textrm{t}$ equals the turnover time $\tau_0$ (see
Eq.~(\ref{eq:m1phi})):
\begin{equation}\label{eq:b4}
\tau_{\rm{t}} = \tau_0 = \frac{{M_0 }}{{F_0}} = \frac{{ \phi \pi^2
(R^2 - r_0^2 )}}{{2K\Delta H \ln(R/r_0)}}
\end{equation}
The average internal age (or life expectancy) $\tau_\textrm{i}$ and
internal transit time $\tau_\textrm{it} = 2\tau_\textrm{i}$ vary
linearly with $\alpha_L$:
\begin{equation}\label{eq:b5}
\tau_{\rm{i}} = \mu_1 [\psi ] = \frac{\phi\pi}{{12 K\Delta H (R^2 -
r_0^2 )}}[8(R^3 - r_0^3)\alpha_L + 3\pi (R^4 - r_0^4)]
\end{equation}
The variance of $\varphi(t)$ can then be deduced from
Eqs.~(\ref{eq:b4}) and~(\ref{eq:b5}) by enforcing
Eq.~(\ref{eq:m1psi}).

\end{document}